\UseRawInputEncoding
\documentclass[showpacs,preprintnumbers,
amsmath,amssymb,aps,prd]{revtex4}
\usepackage{amsfonts}
\usepackage{amsfonts,graphics,epsfig,subfigure}
\usepackage{graphicx}
\usepackage{float}
\begin{document}

\title{Dynamic phase transition of charged dilaton black holes}
\author{Jie-Xiong Mo, Shan-Quan Lan\footnote{shanquanlan@126.com(corresponding
author)}}
\affiliation{Institute of Theoretical Physics, Lingnan Normal University, Zhanjiang, 524048, Guangdong, China}

\begin{abstract}
Dynamic phase transition of charged dilaton black holes is investigated in this paper. We introduce the Gibbs free energy landscape and calculate the corresponding $G_L$ for the dilaton black hole. On the one hand, we numerically solve the Fokker-Planck equation constrained by only the reflecting boundary condition. The effects of dilaton gravity on the probabilistic evolution of dilaton black holes are quite obvious. Firstly, the horizon radius difference between the large dilaton black hole and small dilaton black hole increases with the parameter $\alpha$. Secondly, with the increasing of $\alpha$, the system needs much more time to achieve a stationary distribution. Thirdly, the value which $\rho(r_l,t)$ and $\rho(r_s,t)$ finally attain varies with the parameter $\alpha$. On the other hand, resolving Fokker-Planck equation constrained by both the reflecting boundary condition and absorbing boundary condition, we investigate the first passage process of dilaton black holes. The initial peak decays more slowly with the increase of $\alpha$, which can also be witnessed from the slow down of the decay of $\Sigma(t)$ (the sum of the probability that the black hole system having not finished a first passage by time $t$). Moreover, the time corresponding to the single peak of the first passage time distribution is also found to increase with the parameter $\alpha$. Considering all the observations mentioned above, the dilaton field does slow down the dynamic phase transition process between the large black hole and small black hole.
\end{abstract}
   \maketitle

\section{Introduction}

     Dilaton gravity~\cite{Gibbons1,Garfinkle,Koikawa} has received considerable attention within the community for its great physical significance. Firstly, it serves as the low energy limit of string theory. One can recover Einstein gravity with a non-minimally coupled dilaton field and other fields in the low energy limit. Secondly, there exist one or more Liouville-type potentials in its action. And these potentials can be caused by spacetime supersymmetry breaking in ten dimensions. Thirdly, it has been disclosed that the dilaton field may affect both the casual structure and thermodynamic properties of black holes. Due to the above physical importance, various black hole solutions were presented with their thermodynamic properties discussed~\cite{Gibbons1}-~\cite{zhangchengyong}. Interestingly, it was argued in Refs.~\cite{gaochangjun1,gaochangjun2,gaochangjun3} that the cosmological constant is found to be coupled to Liouville-type potential in the dilaton theory.

     Among these solutions, a class of $(n+1)$-dimensional topological black hole solutions were obtained~\cite{Sheykhi2} in Einstein-Maxwell-dilaton theory, which are neither asymptotically flat nor anti-de Sitter. Thermodynamic properties of these black holes were carefully analyzed. Moreover, the effect of dilaton field on thermal stability was disclosed~\cite{Sheykhi2}. $P-V$ criticality was investigated in Refs.~\cite{zhaoren,Sheykhi1}. Provided the coupling constant of dilaton gravity is less than one, the critical values of relevant physical quantities are physical~\cite{Sheykhi1}. Ref.~\cite{Sheykhi13} probed the phase transition and thermodynamic geometry while Ref.~\cite{Sheykhi16} further disclosed the existence of zeroth-order phase transition. Both the two point correlation function and the entanglement entropy of dilaton black holes were studied by one of us~\cite{xiong10}, providing an alternative perspective to observe the critical phenomena in dilaton black holes. With the tool of Ruppeiner geometry, Ref.~\cite{zhangshaojun} tried to provide the microscopic explanation for phase transitions of charged dilatonic black holes.

     Although many attempts have been made as stated above, the dynamical phase transition of this typical class of dilaton black hole solution proposed in Ref.~\cite{Sheykhi2} and the possible kinetics behind it remains to be probed. So it is the main motivation of this paper to complete this ``missing puzzle piece" concerning the phase transition of charged dilaton black holes. Dynamic phase transition of black holes has been a newly emerged hot topic ever since the creative works reporting the kinetics of the Hawing-Page phase transition~\cite{liran1} and the van der Waals like phase transition~\cite{liran2} recently. For the latter, the transition between the large black hole and the small black hole due to thermal fluctuation can be viewed as stochastic process, which can be probed via the tool of Fokker-Planck equation~\cite{liran2}. Soon after, dynamic phase transition of Gauss-Bonnet black holes~\cite{liran3,weishaowen2,weishaowen3} and charged AdS black holes surrounded by quinteness dark energy~\cite{xiong11} were investigated. More interestingly, it was even suggested in a very recent work~\cite{liran4} that the turnover of the kinetics can be used to probe the microstructure of black holes. It is certainly of interest to generalize these pioneer works to dilaton black holes considering the great physical importance of dilaton gravity stated in the first paragraph of this paper. It is expected that novel features will be disclosed about the dynamic phase transition of black holes.

    The organization of this paper is as follows. Sec.\ref{Sec2} is devoted to a short review of thermodynamic properties of charged dilaton black holes. In Sec.\ref{Sec3}, we will investigate the dynamic phase transition of charged dilaton black holes. We will further probe the first passage process for the phase transition of charged dilaton black holes in Sec.\ref{Sec4}. Conclusions will be presented in Sec.\ref{Sec5}.

\section{A short review on thermodynamic properties of charged dilaton black holes}
\label{Sec2}

 The Einstein-Maxwell-Dilaton action in $(n+1)$-dimensional spacetime reads~\cite{Chan}
\begin{equation}
S=\frac{1}{16\pi}\int d^{n+1}x \sqrt{-g}\left(\mathcal{R}-\frac{4}{n-1}(\nabla \Phi)^2-V(\Phi)-(e^{-4\alpha \Phi/(n-1)}F_{\mu\nu}F^{\mu\nu}\right),\label{1}\\
\end{equation}
where  $\mathcal{R}$, $\Phi$ and $F_{\mu\nu}$ denote respectively the Ricci scalar curvature, the dilaton field and the electromagnetic field tensor. The parameter $\alpha$ describes the strength of coupling between the electromagnetic field and the dilaton field.

To obtain a solution, one needs to take a specific form for the potential of the dilaton field $V(\Phi)$. Here, the potential of the dilaton field is chosen as $V(\Phi)=2\Lambda_0e^{2\zeta_0\Phi}+2\Lambda e^{2\zeta\Phi}$~\cite{Chan,Sheykhi4,Sheykhi2,Sheykhi3}. Note that if one consider the power-law Maxwell field~\cite{Sheykhi8}, an additional term should be added to the potential mentioned above.

One can take the general form of metric as~\cite{Sheykhi2}
\begin{equation}
ds^2=-f(r)dt^2+\frac{1}{f(r)}dr^2+r^2R^2(r)h_{ij}dx^idx^j,\label{2}\\
\end{equation}
where $h_{ij}dx^idx^j$ corresponds to $(n-1)$-dimensional hypersurface with constant scalar curvature $(n-1)(n-2)k$. $k$ can be taken as $-1,\,0,\,1$, corresponding to hyperbolic, flat and spherical constant curvature hypersurface respectively.

Making the ansatz $R=e^{2\alpha\Phi/(n-1)}$, the corresponding solution was derived~\cite{Sheykhi2} as
\begin{eqnarray}
f(r)&=&-\frac{k(n-2)(1+\alpha^2)^2b^{-2\gamma}r^{2\gamma}}{(\alpha^2-1)(\alpha^2+n-2)}-\frac{m}{r^{(n-1)(1-\gamma)-1}}+\frac{2\Lambda b^{2\gamma}(1+\alpha^2)^2r^{2(1-\gamma)}}{(n-1)(\alpha^2-n)}
\nonumber
\\
&\,&+\frac{2q^2(1+\alpha^2)^2b^{-2(n-2)\gamma}r^{2(n-2)(\gamma-1)}}{(n-1)(n+\alpha^2-2)},\label{3}\\
\Phi(r)&=&\frac{(n-1)\alpha}{2(\alpha^2+1)}\ln \left(\frac{b}{r}\right),\label{4}
\end{eqnarray}
where $b$ is an arbitrary constant and $\gamma$ is related to $\alpha$ by $\gamma=\alpha^2/(\alpha^2+1)$. And the coefficients in the expression of $V(\Phi)$ read~\cite{Sheykhi2}
\begin{equation}
\Lambda_0=\frac{k(n-1)(n-2)\alpha^2}{2b^2(\alpha^2-1)},\;\zeta_0=\frac{2}{\alpha(n-1)},\;\zeta=\frac{2\alpha}{n-1}.\label{5}\\
\end{equation}

The mass, the charge, Hawking temperature and the entropy was obtained respectively as~\cite{Sheykhi2,zhaoren}
\begin{eqnarray}
M&=&\frac{b^{(n-1)\gamma}(n-1)\omega_{n-1}m}{16\pi (\alpha^2+1)} ,\label{6}
\\
Q&=&\frac{\omega_{n-1}q}{4\pi},\label{7}\\
T&=&\frac{-(1+\alpha^2)}{2\pi(n-1)}\Big[\frac{k(n-2)(n-1)}{2b^{2\gamma}(\alpha^2-1)}r_+^{2\gamma-1}+\Lambda b^{2\gamma} r_+^{1-2\gamma}+q^{2}b^{-2(n-2)\gamma}r_+^{(2n-3)(\gamma-1)-\gamma}\Big],\label{8}\\
S&=&\frac{\omega_{n-1}b^{(n-1)\gamma}r_+^{(n-1)(1-\gamma)}}{4}.\label{9}
\end{eqnarray}
where the parameter $m$ can be derived via $f(r_+)=0$.

Considering the demand that the electric potential $A_t$ should be finite at infinity and the parameter $m$ should vanish at spacial infinity, restrictions on $\alpha$ for dilaton black holes coupled with power-law Maxwell field was obtained~\cite{Sheykhi8} as $0\leq\alpha^2<n-2$ for the case $\frac{1}{2}<p<\frac{n}{2}$ while $2p-n<\alpha^2<n-2$ for the case $\frac{n}{2}<p<n-1$, where $p$ describes the nonlinearity of the electromagnetic field. For the Maxwell field here, $p=1$. And the corresponding restriction reduces to be $0\leq\alpha^2<n-2$.

\section{Dynamic phase transition of charged dilaton black holes}
\label{Sec3}

To probe the dynamic phase transition of black holes, one should introduce the ``Gibbs free energy landscape" and define the corresponding Gibbs free energy $G_L$ as $G_L=H-T_ES$. Here, $T_E$ denotes the temperature of the ensemble. As suggested in the pioneer work~\cite{liran1,liran2,weishaowen2}, $G_L$ possesses the following properties. Firstly, only when $T_E=T$, $G_L$ describes a real black hole. Secondly, only the extremal points of $G_L$ make sense. The local maximum and minimum denote respectively the unstable and stable black hole phases. Thirdly, the small-large black hole phase transition corresponds to the double wells of $G_L$ having the same depth.

Via Eqs. (\ref{6}) and (\ref{9}), $G_L$ of the dilaton black hole can be calculated as
\begin{eqnarray}
G_L&=&\frac{\omega_{n-1}b^{(n-1)\gamma}r_+^{n-2+(1-n)\gamma}}{16\pi}\Big\{\Big[\frac{2b^{-2(n-2)\gamma} q^2r_+^{2(n-2)(\gamma-1)}}{(n-1)(\alpha^2+n-2)}-\frac{k(n-2)b^{-2\gamma}r^{2\gamma}}{(\alpha^2-1)(\alpha^2+n-2)}+\frac{16\pi P b^{2\gamma}r^{2(1-\gamma)}}{(n-1)(n-\alpha^2)}\Big] \nonumber
\\
&\;&\times(n-1)(1+\alpha^2)-4\pi r_+T_E\Big\},\label{10}
\end{eqnarray}
where we have introduced the extended phase space and identify the thermodynamic pressure as $P=-\Lambda/8\pi$.

Based on $G_L$, we can utilize the following Fokker-Planck equation~\cite{Zwanzig,Lee1,Lee2,wang1,Bryngelson} to investigate the probabilistic evolution of the black hole after the thermal fluctuation
\begin{equation}
\frac{\partial \rho(r,t)}{\partial t}=D\frac{\partial}{\partial r}\left\{e^{-\beta G_L(T,P,r)}\frac{\partial}{\partial r}\left[e^{\beta G_L(T,P,r)}\rho(r,t)\right]\right\},\label{11}
\end{equation}
where $\rho(r,t)$ denotes the probability distribution. The parameter $\beta=\frac{1}{k_B T}$ with $k_B$ as the Boltzman constant. The diffusion coefficient $D=\frac{k_B T}{\zeta}$ with $\zeta$ denoting the dissipation coefficient. Without loss of generality, one can set $\zeta$ and $k_B$ as one. For the sake of conciseness, we use the notation $r$ to replace the notation $r_+$ for the black hole horizon radius here and hereafter.

Before solving the above equation, one should impose both the boundary condition and the initial condition first. The boundary conditions can be divided into two categories, reflecting boundary condition and absorbing boundary condition. As suggested in Ref.~\cite{liran2}, the reflecting boundary condition can be considered as $\beta G'(r)\rho(r,t)+\rho'(r,t)\big|_{r=r_0}=0$ while absorbing boundary condition can be considered as $\rho(r_m,t)=0$ (The meaning of $r_m$ would be discussed in the next paragraph). The initial condition can be chosen as Gaussian wave pocket located at the small/large dilaton black hole state respectively. Namely, $\rho(r,0)=\frac{1}{\sqrt{\pi}a}e^{-(r-r_{l(s)})^2/a^2}$, with $r_l$ and $r_s$ denote the horizon radius of large dilaton black hole and small dilaton black hole respectively~\cite{liran2}.

\begin{figure}[H]
\centerline{\subfigure[]{\label{a0.4p0.003l1}
\includegraphics[width=8cm,height=6cm]{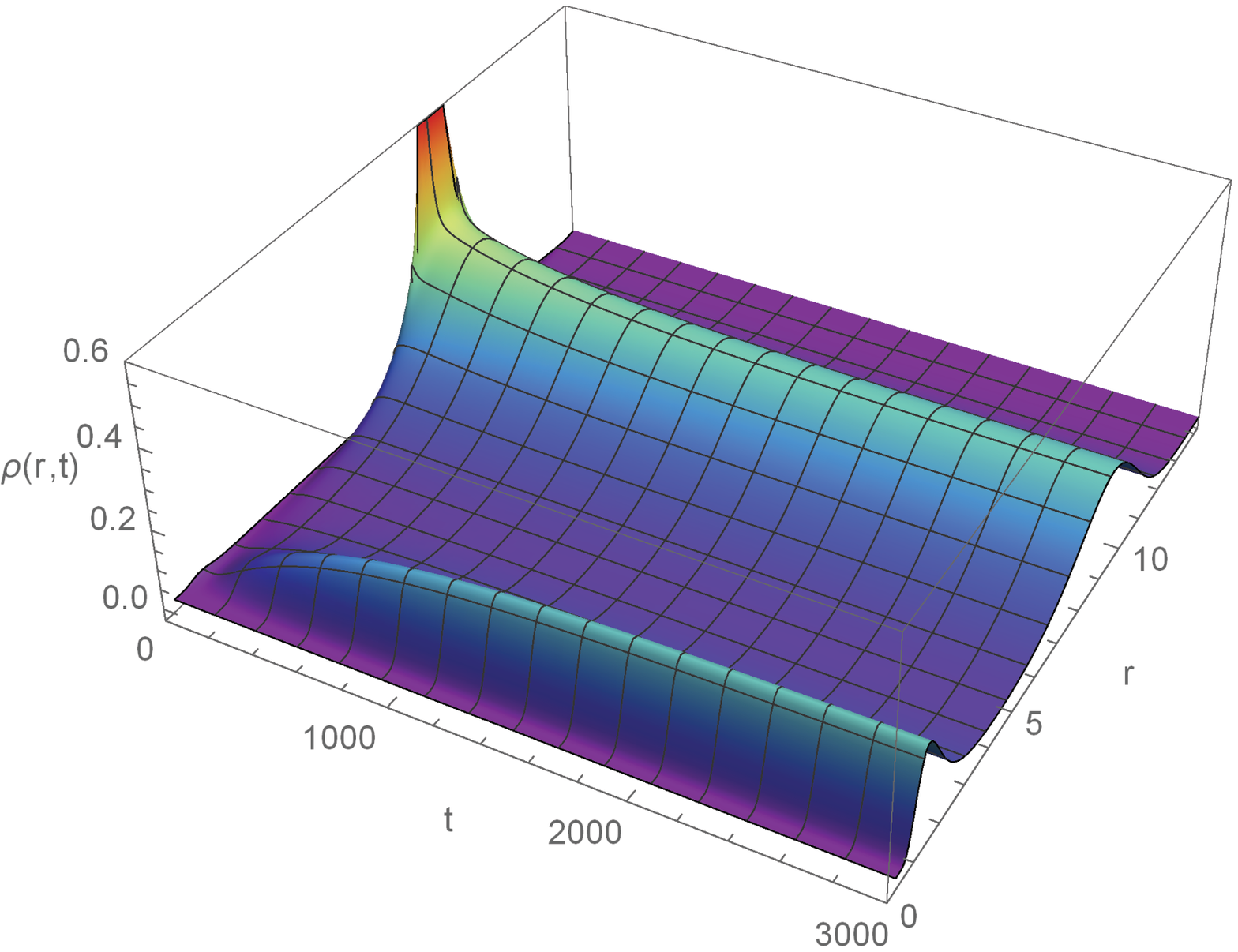}}
\subfigure[]{\label{a0.4p0.003s1}
\includegraphics[width=8cm,height=6cm]{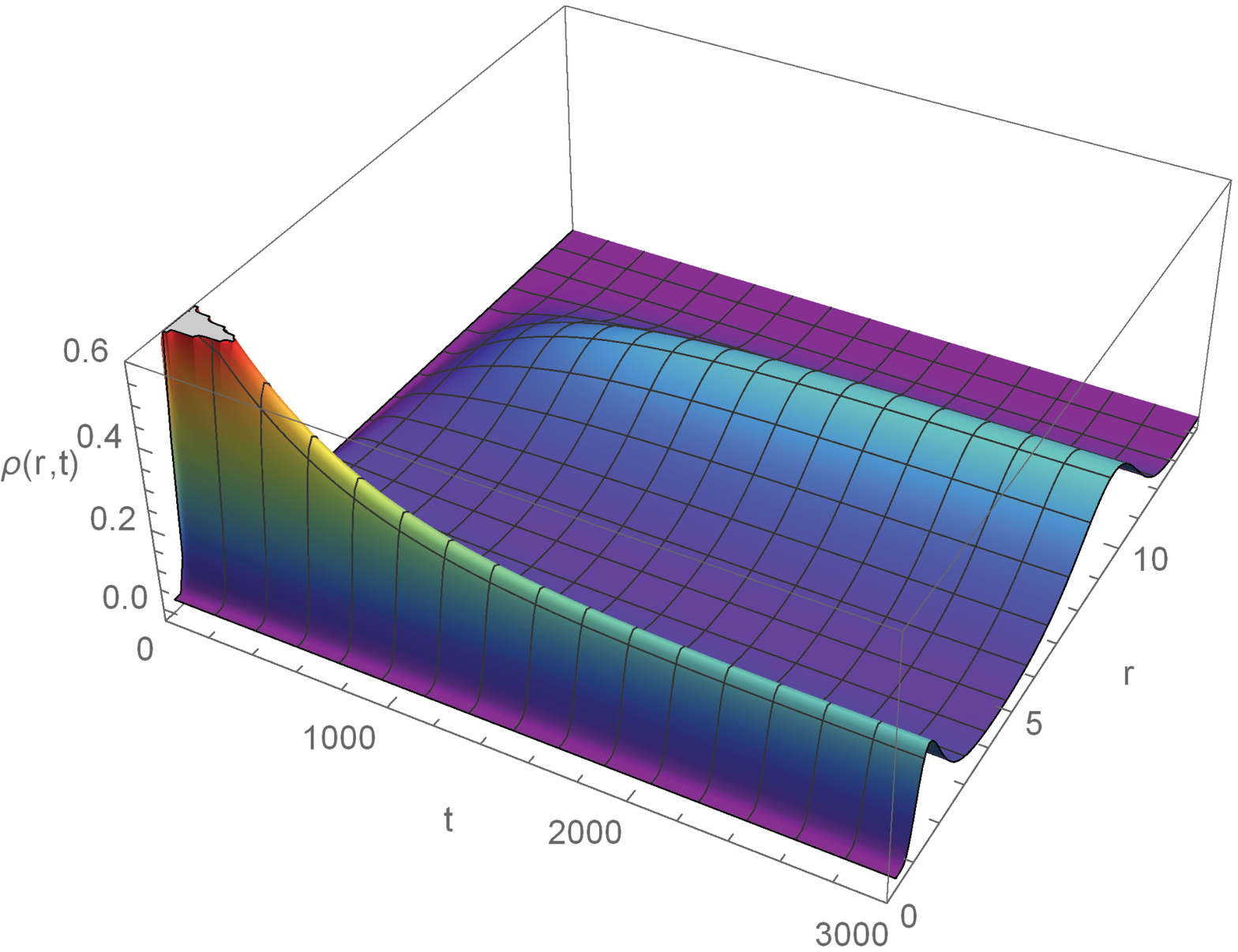}}}
\centerline{\subfigure[]{\label{a0.2p0.003l1}
\includegraphics[width=8cm,height=6cm]{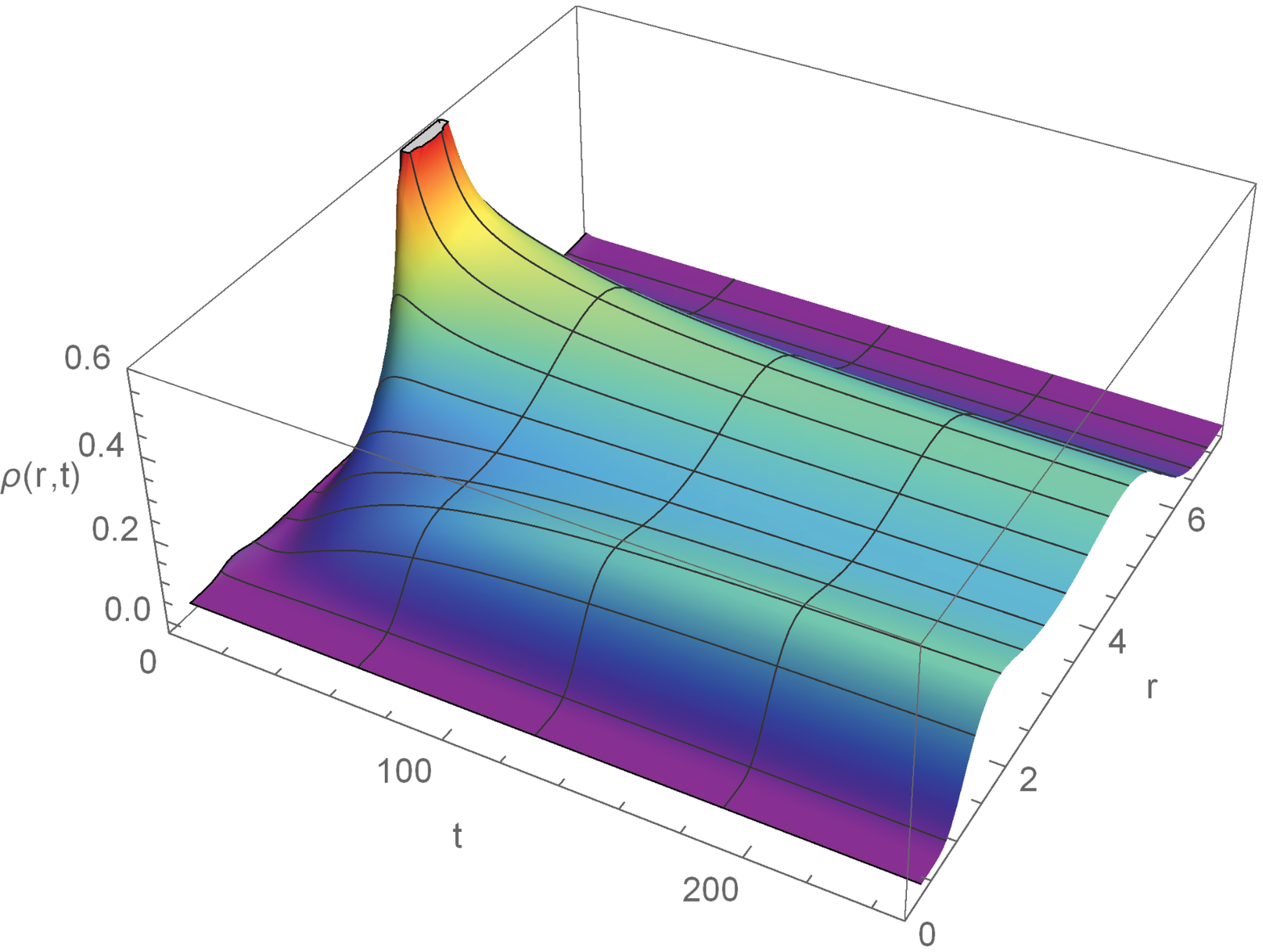}}
\subfigure[]{\label{a0.2p0.003s1}
\includegraphics[width=8cm,height=6cm]{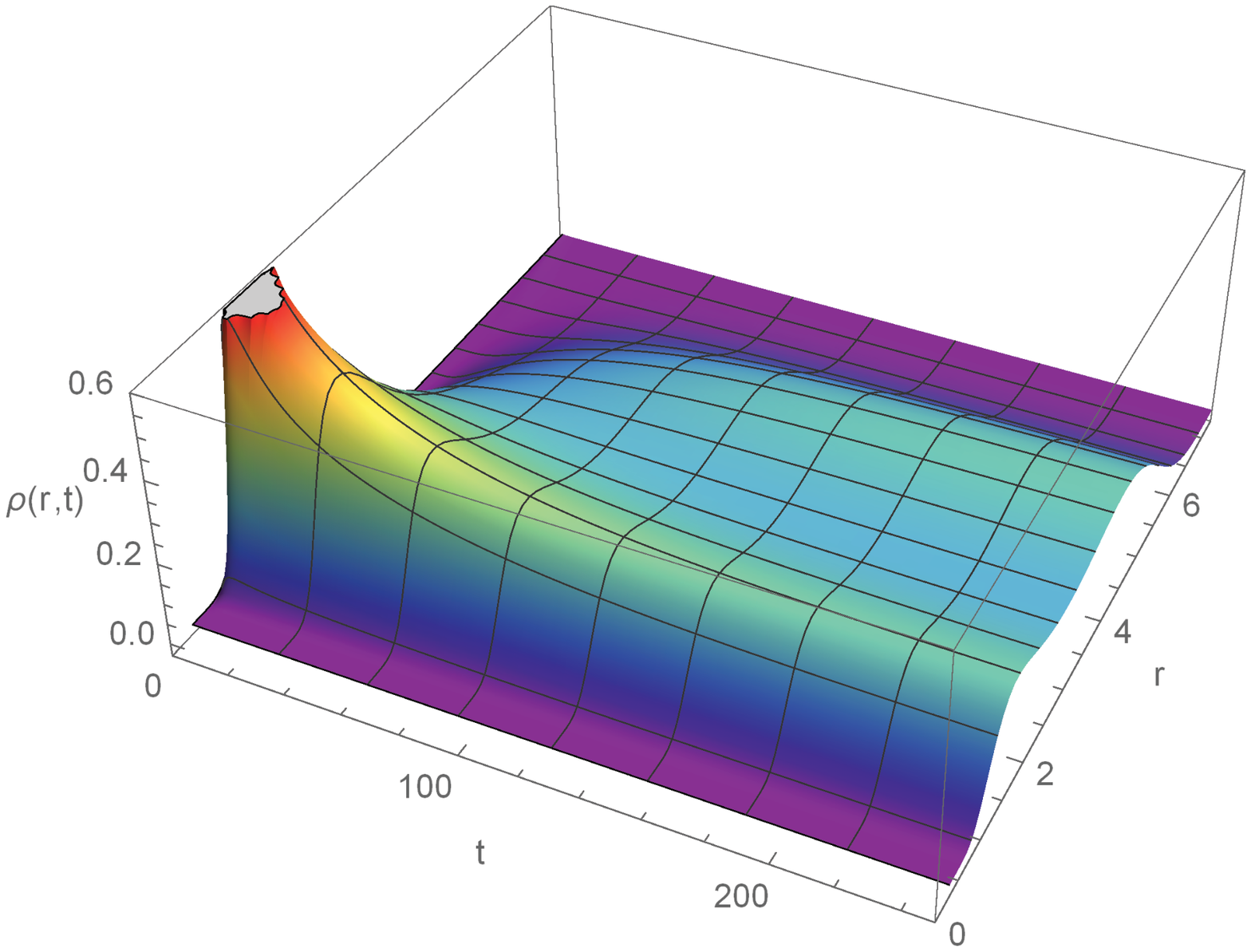}}}
\centerline{\subfigure[]{\label{a0p0.003l1}
\includegraphics[width=8cm,height=6cm]{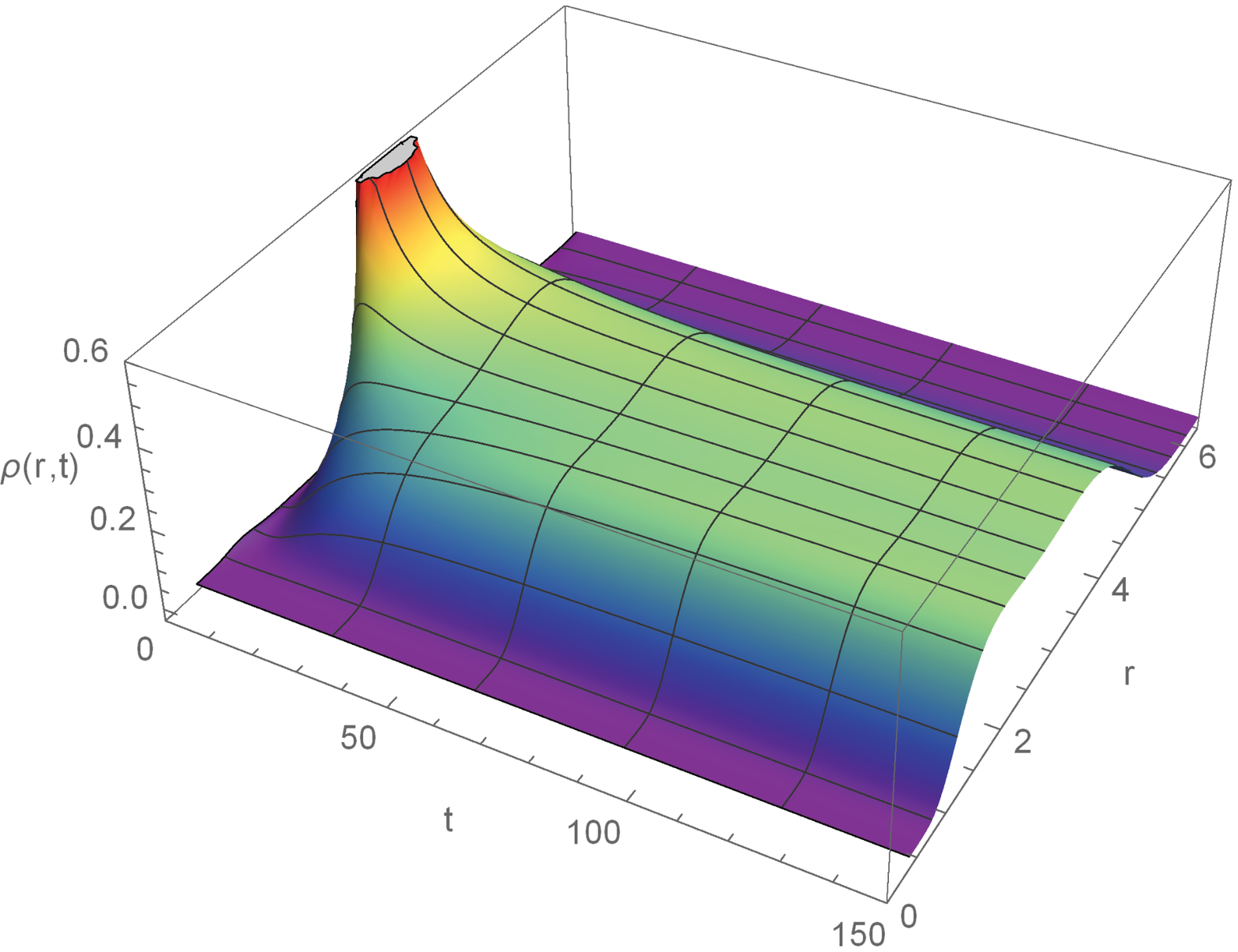}}
\subfigure[]{\label{a0p0.003s1}
\includegraphics[width=8cm,height=6cm]{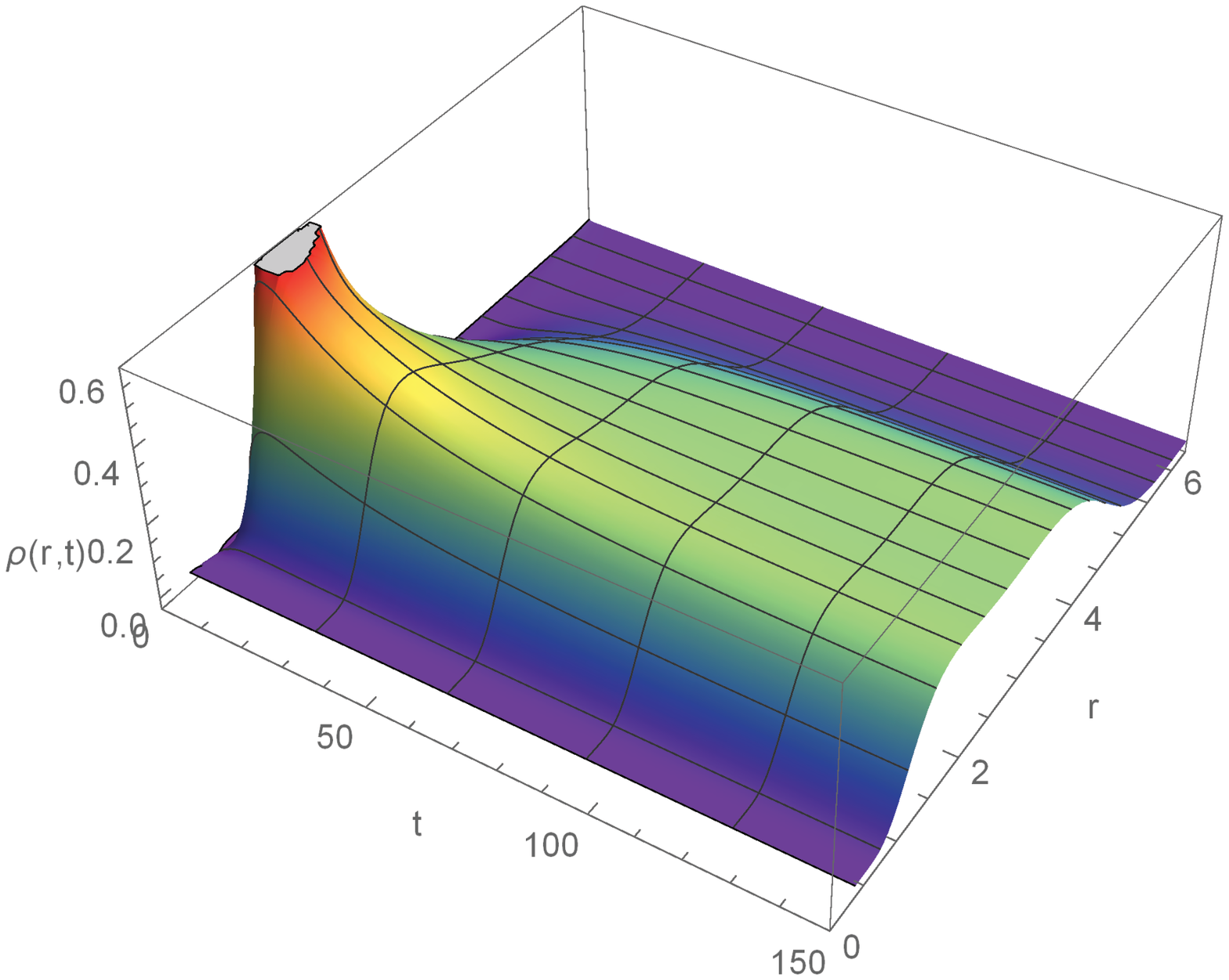}}}
 \caption{Time evolution of $\rho(r,t)$ for charged dilaton black hole with $b=1, q=1, P=0.003$. (a) $\alpha=0.4$ (b) $\alpha=0.4$  (c) $\alpha=0.2$  (d) $\alpha=0.2$ (e) $\alpha=0$ (f) $\alpha=0$. For the left three graphs, the initial condition is chosen as Gaussian wave pocket located at the large dilaton black hole state while for the right three graphs, the initial condition is chosen as Gaussian wave pocket located at the small dilaton black hole state.}
\label{fg1}
\end{figure}

We numerically solve Eq.(\ref{11}) constrained by both the initial condition and the reflecting boundary condition mentioned above. Fig.\ref{fg1} shows the time evolution of the probability distribution $\rho(r,t)$ of charged dilaton AdS black holes, where the thermodynamic pressure is chosen as $P=0.003$. In the top, middle and bottom rows of Fig.\ref{fg1}, the parameter characterizing the strength of coupling between the electromagnetic field and the dilaton field $\alpha$ is chosen to be $0.4,0.2,0$ respectively to compare the effect of dilaon gravity on the probability evolution.

For the left three graphs of Fig.\ref{fg1}, the initial condition is chosen as Gaussian wave pocket located at the large dilaton black hole state. This can be reflected in the graphs that the initial peak of $\rho(r,t)$ corresponds to the radius of the large dilaton black hole. With the time evolution, the initial peak is decreasing while another peak corresponds to the radius of the small dilaton black hole is booming. Finally, these two peaks reach a stationary distribution that they equal to each other at the first sight. This clearly exhibits the dynamic process of how the initial large dilaton black hole evolves into the small dilaton black hole. So the left three graphs correspond to the case the large dilaton black hole evolves dynamically into small dilaton black hole.

By contrast, the initial condition is chosen as Gaussian wave pocket located at the small dilaton black hole state for the right three graphs of Fig.\ref{fg1}. The initial peak locating at the small dilaton black hole is decaying while another peak corresponding to the radius of the large dilaton black hole is increasing with the time evolution. At the long time limit, these two peaks corresponding to the large (small) black holes respectively approach to a stationary distribution. So the right three graphs of Fig.\ref{fg1} correspond to the case the small dilaton black hole evolves dynamically into large dilaton black hole.

%%%%%%%%%%%%%%%%%%%%%%%%%%%%%%%%%%%%%%%%%%%%%%%%%%%%%%%%%%%%%%%%%%%%%%%%%%%%%
%%%%%%%%%%%%%%%%%%%%%%%%%%%%%%%%%%%%%%%%%%%%%%%%%%%%%%%%%%%%%%%%%%%%%%%%%%%%%%%%

%%%%%%%%%%%%%%%%%%%%%%%%%%%%%%%%%%%%%%%%%%%%%%%%%%%%%%%%%%%%%%%%%%%%%%%%%%%%%
\begin{figure}[H]
\centerline{\subfigure[]{\label{a0.4p0.003l2}
\includegraphics[width=8cm,height=6cm]{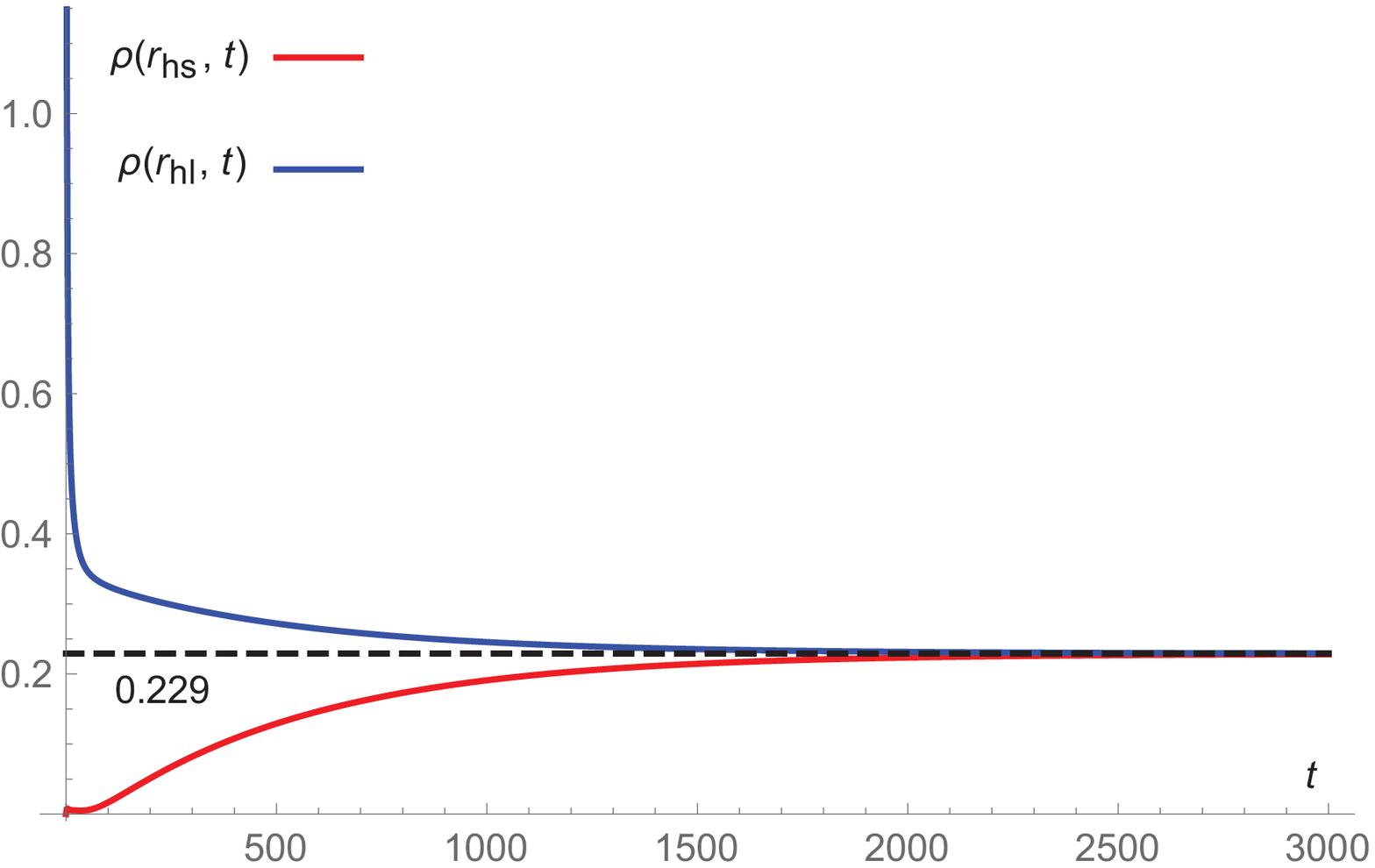}}
\subfigure[]{\label{a0.4p0.003s2}
\includegraphics[width=8cm,height=6cm]{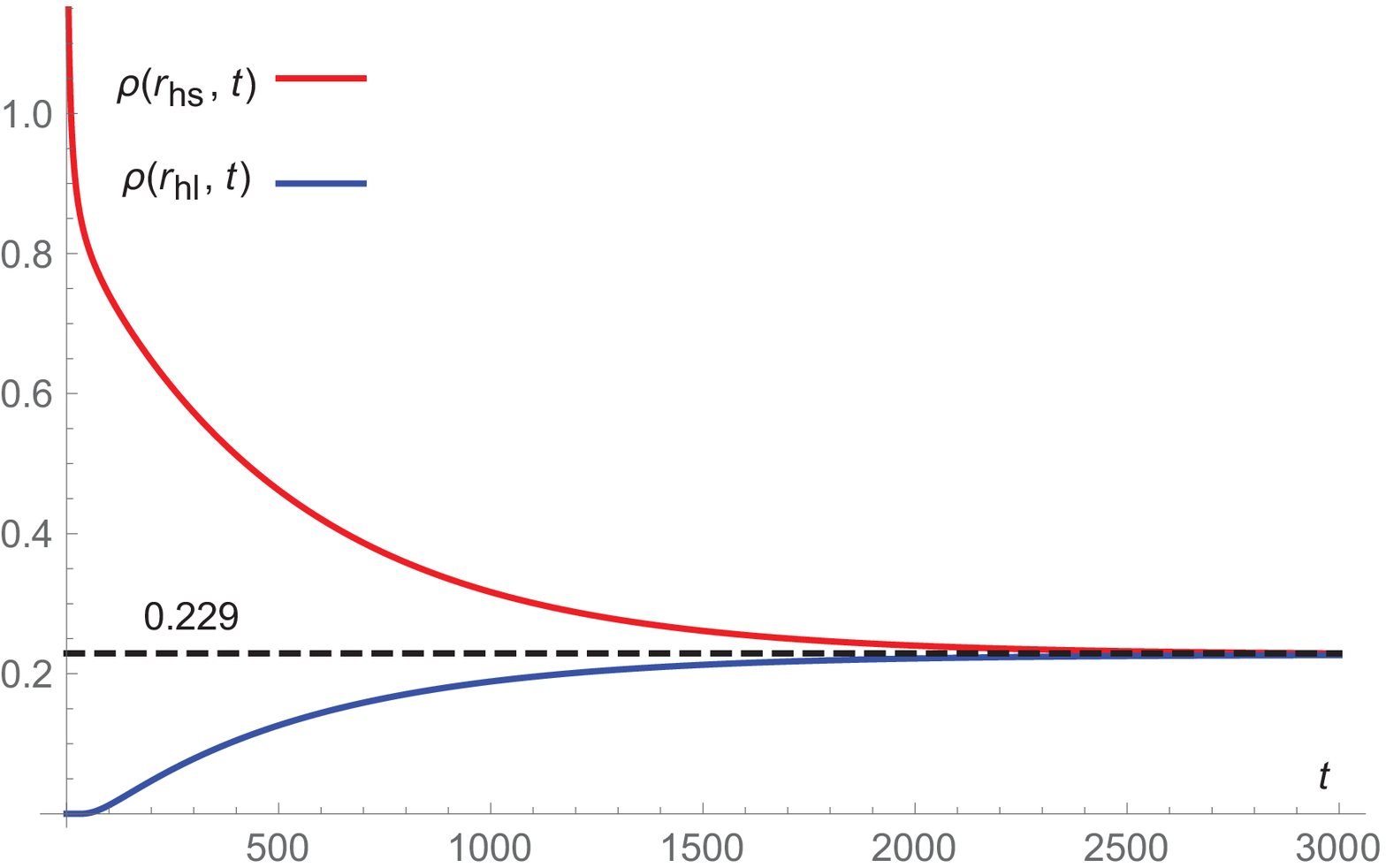}}}
\centerline{\subfigure[]{\label{a0.2p0.003l2}
\includegraphics[width=8cm,height=6cm]{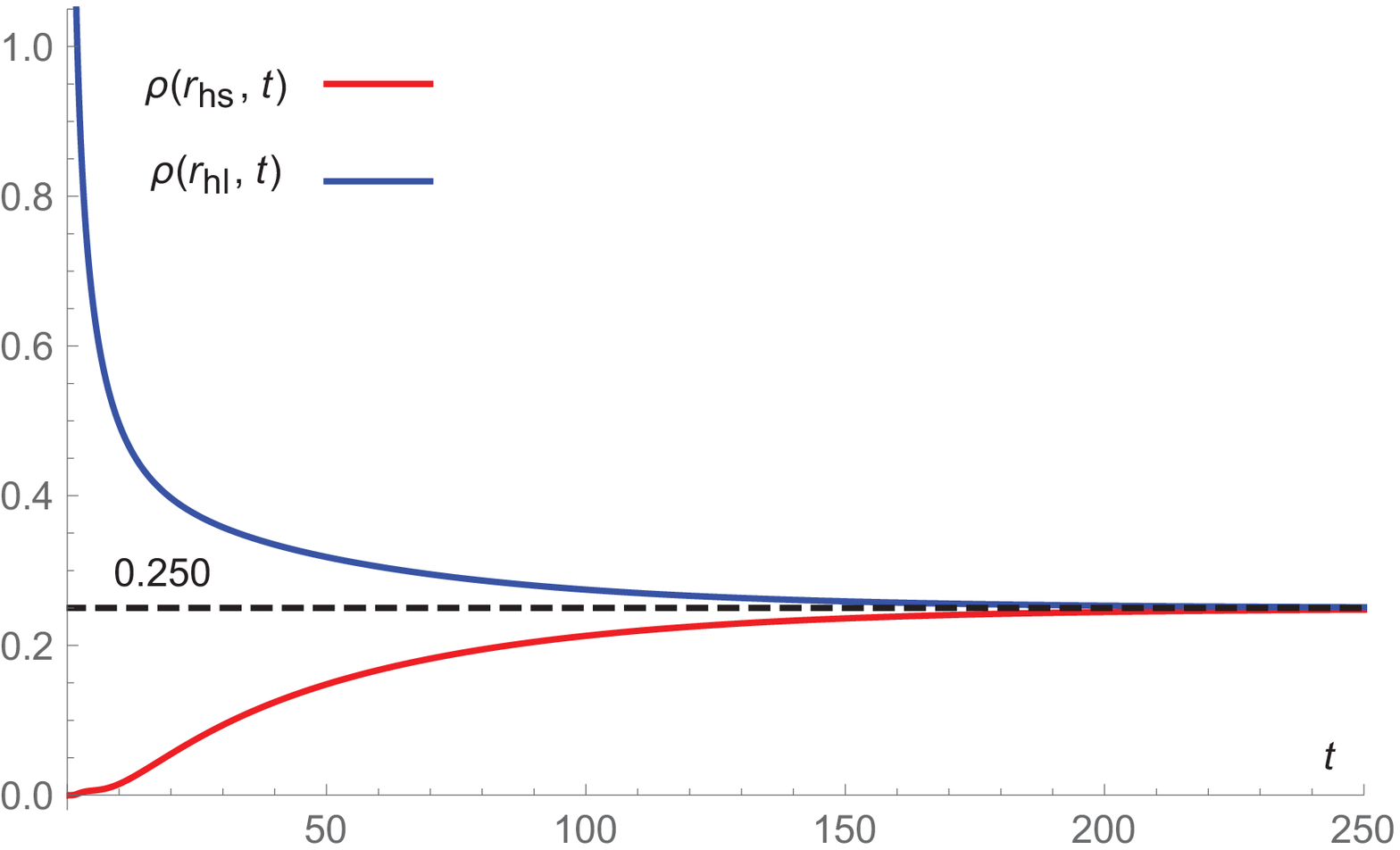}}
\subfigure[]{\label{a0.2p0.003s2}
\includegraphics[width=8cm,height=6cm]{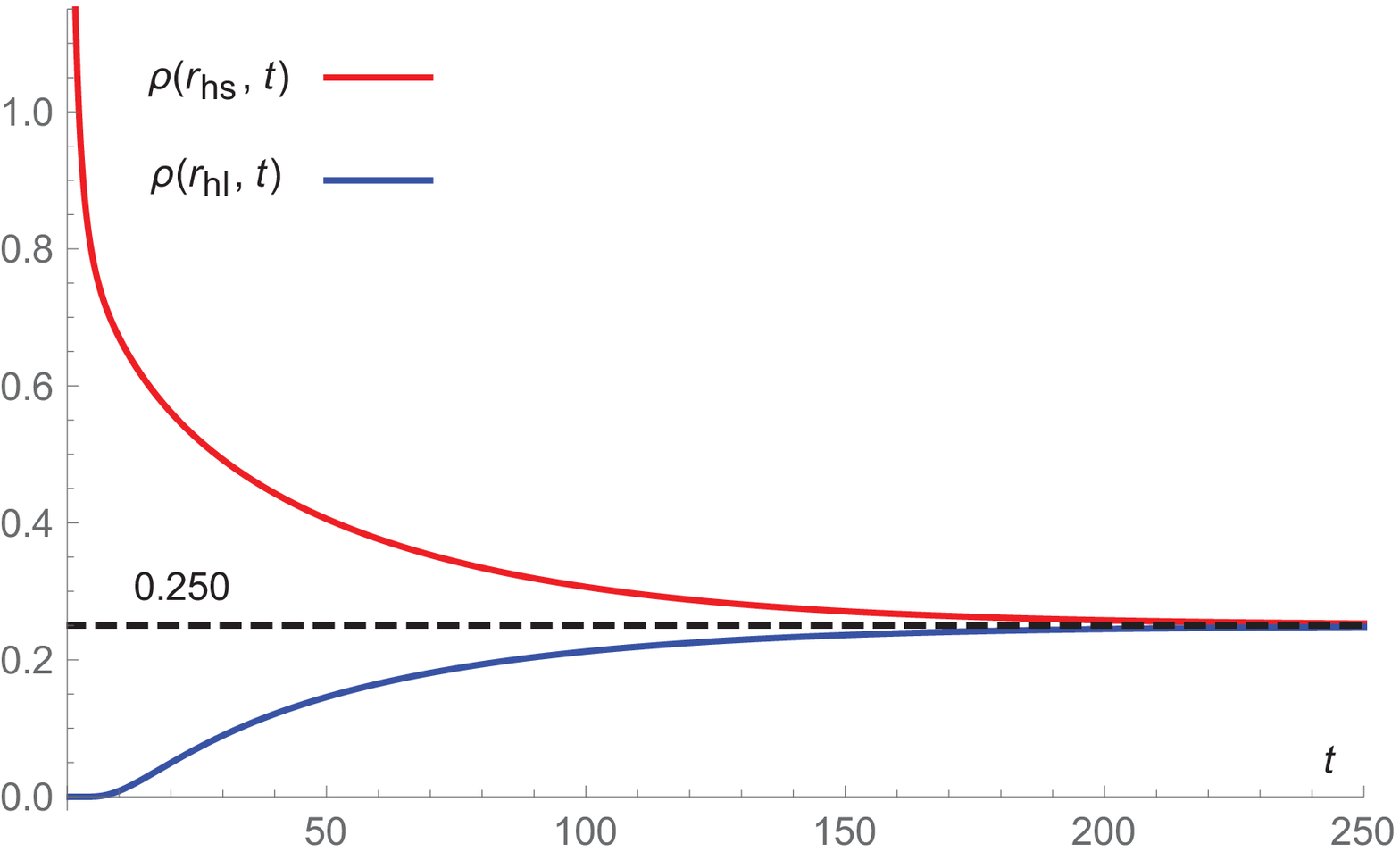}}}
\centerline{\subfigure[]{\label{a0p0.003l2}
\includegraphics[width=8cm,height=6cm]{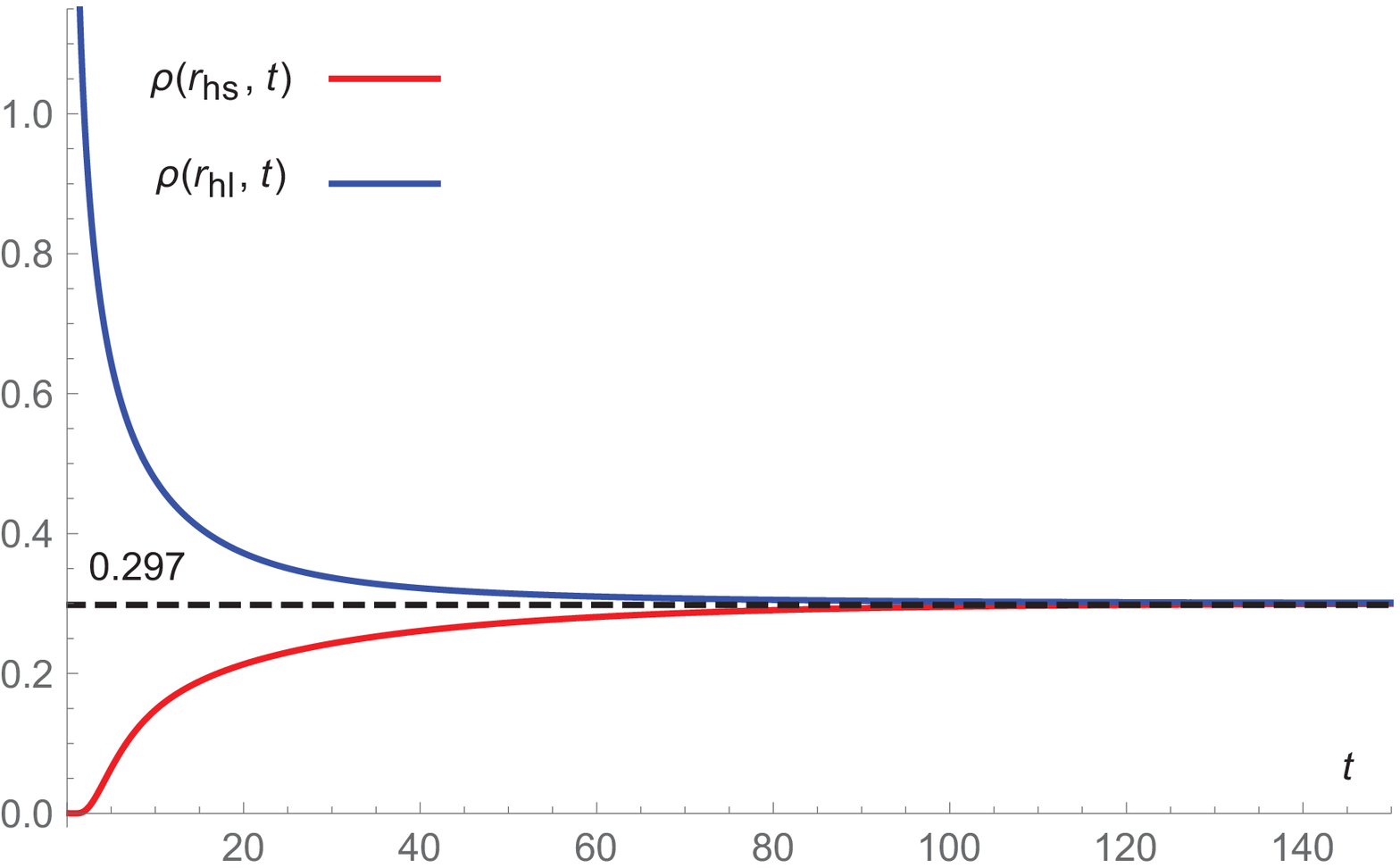}}
\subfigure[]{\label{a0p0.003s2}
\includegraphics[width=8cm,height=6cm]{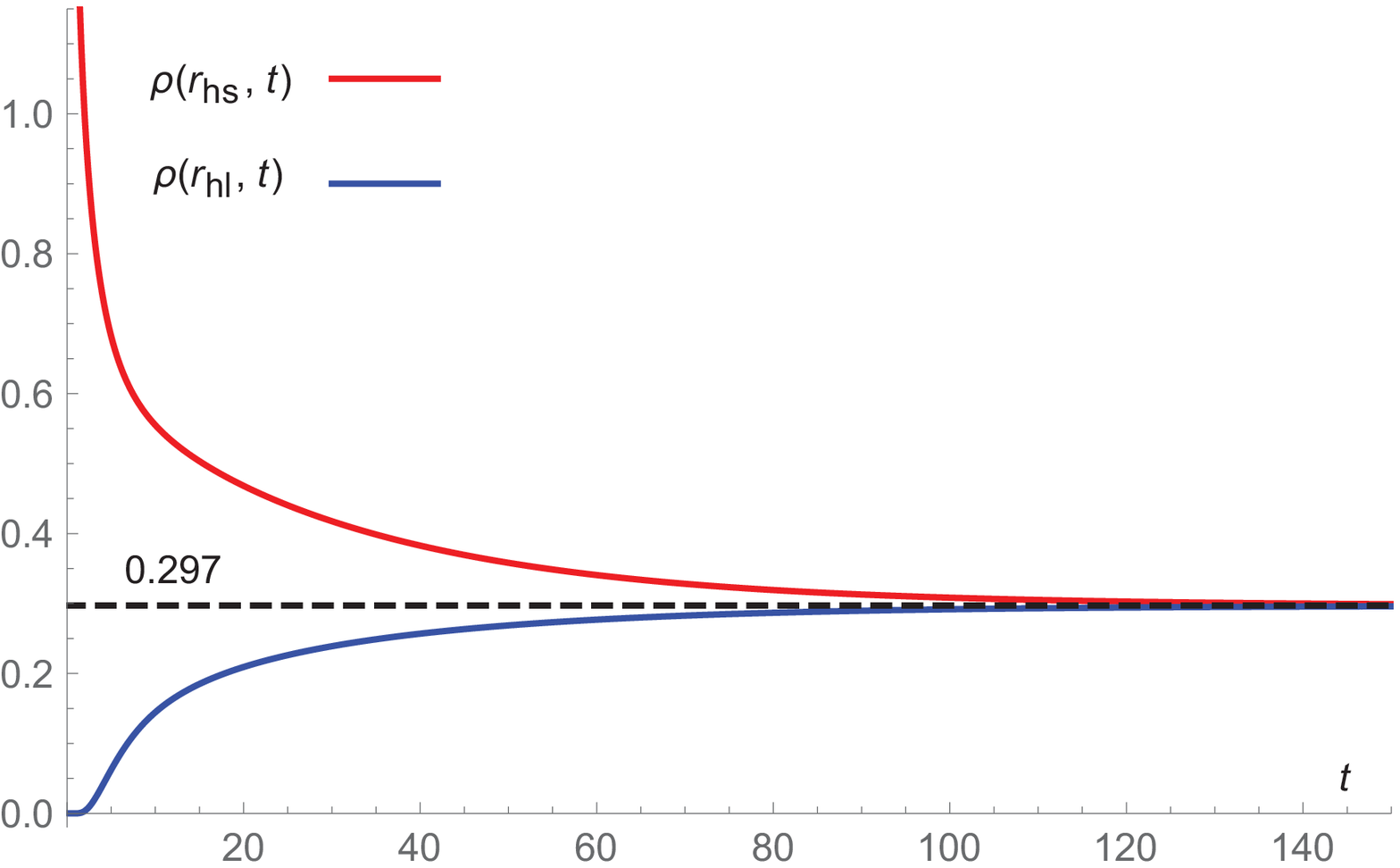}}}
 \caption{Time evolution of $\rho(r_l,t)$ and $\rho(r_s,t)$ for charged dilaton black hole with $b=1, q=1, P=0.003$. (a) $\alpha=0.4$ (b) $\alpha=0.4$  (c) $\alpha=0.2$  (d) $\alpha=0.2$ (e) $\alpha=0$ (f) $\alpha=0$. For the left three graphs, the initial condition is chosen as Gaussian wave pocket located at the large dilaton black hole state while for the right three graphs, the initial condition is chosen as Gaussian wave pocket located at the small dilaton black hole state.}
\label{fg2}
\end{figure}
%%%%%%%%%%%%%%%%%%%%%%%%%%%%%%%%%%%%%%%%%%%%%%%%%%%%%%%%%%%%%%%%%%%%%%%%%%%%%%%%

Comparing the top two graphs ($\alpha=0.4$), the middle two graphs ($\alpha=0.2$) with the bottom two graphs ($\alpha=0$) in Fig.\ref{fg1}, the effect of dilaton gravity is quite obvious. Firstly, the horizon radius difference between the large dilaton black hole and small dilaton black hole increases with the parameter $\alpha$. Secondly, with the increasing of $\alpha$, the system needs much more time to achieve a stationary distribution. Note that these two observations do not depend on the initial condition.

 Fig.\ref{fg2} provides a complementary but much more quantitative picture of the time evolution of $\rho(r_l,t)$ and $\rho(r_s,t)$. As can be seen from the left three graphs of Fig.\ref{fg2}, $\rho(r_l,t)$ is finite while $\rho(r_s,t)$ equals to zero at first. Then $\rho(r_l,t)$ starts to decrease while $\rho(r_s,t)$ begins to increase. Finally, they attain the same value. The value can be read from the graph, $0.229$ for the case of $\alpha=0.4$, $0.250$ for the case of $\alpha=0.2$ and $0.297$ for the case of $\alpha=0$. This difference can be viewed as the third kind of effect (besides the two observations stated in the former paragraph) of the dilaon gravity on the dynamic phase transition of black holes. The situation for the right three graphs is quite the reverse. At the beginning, $\rho(r_s,t)$ is finite while $\rho(r_l,t)$ equals to zero. $\rho(r_s,t)$ decays with time while $\rho(r_l,t)$ increases with time. In the end, they also reach the same value, as in the left three graphs.

\section{First passage process for the phase transition of charged dilaton black holes}
\label{Sec4}

To further probe the dynamic phase transition of dilaton black holes, one can also consider first passage process. This process describes how the initial black hole state approach the intermediate transition state for the first time. To numerically investigate such a process, both the reflecting boundary condition and absorbing boundary condition should be imposed. Note that absorbing boundary condition $\rho(r_m,t)=0$~\cite{liran2} is considered at the the intermediate transition state whose horizon radius denoted as $r_m$.

\begin{figure}[H]
\centerline{\subfigure[]{\label{a0.4p0.003l3}
\includegraphics[width=8cm,height=6cm]{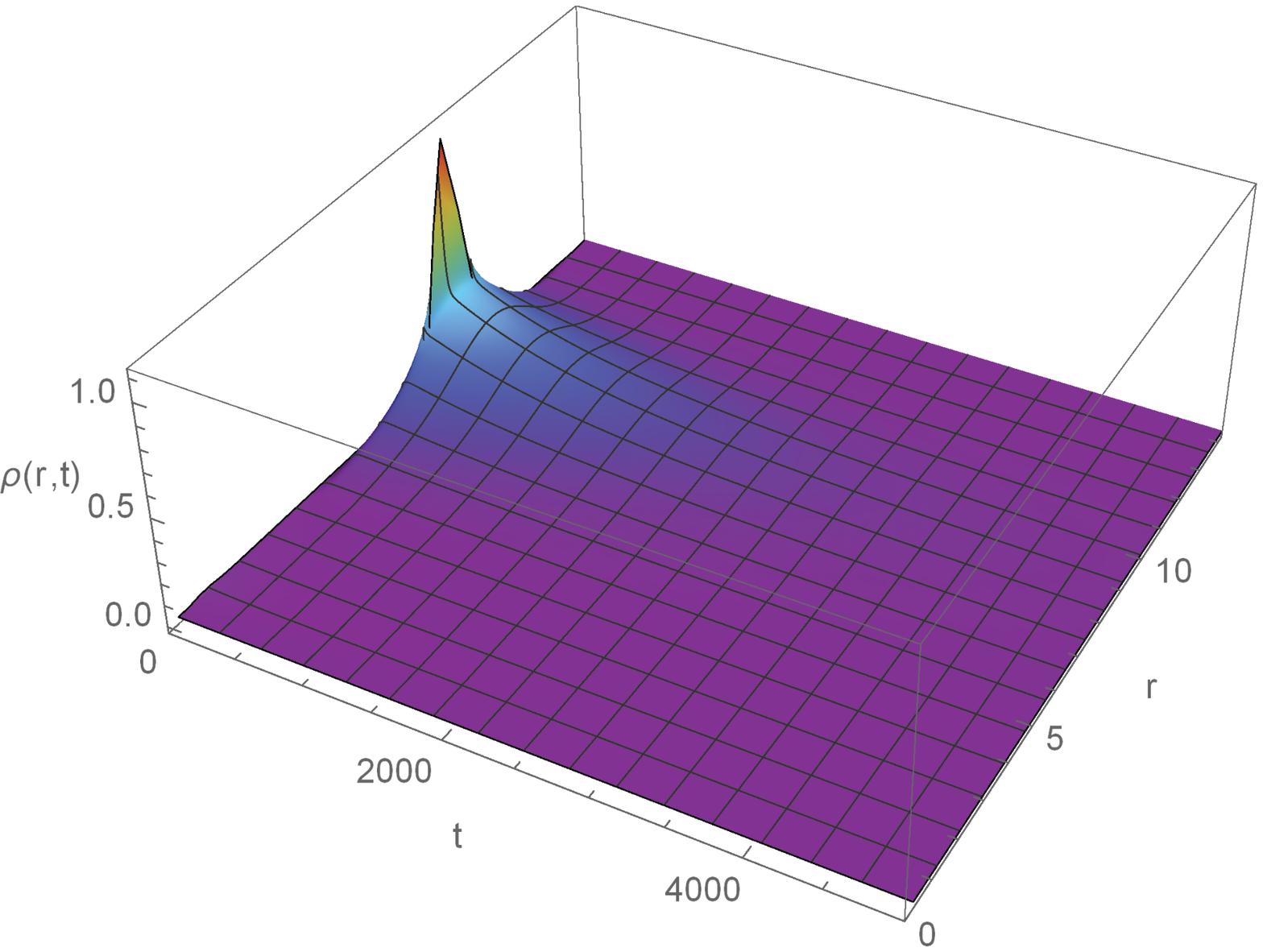}}
\subfigure[]{\label{a0.4p0.003s3}
\includegraphics[width=8cm,height=6cm]{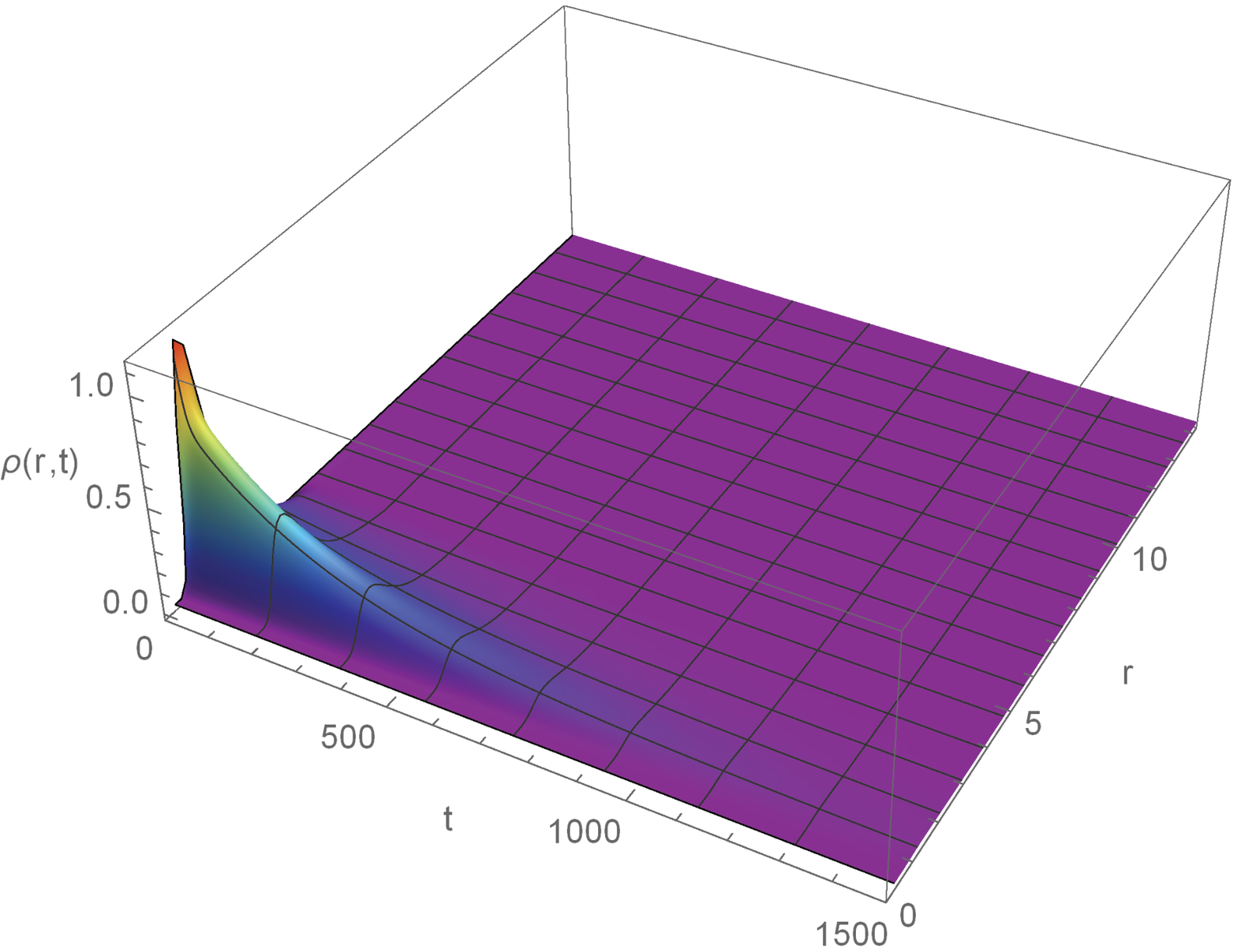}}}
\centerline{\subfigure[]{\label{a0.2p0.003l3}
\includegraphics[width=8cm,height=6cm]{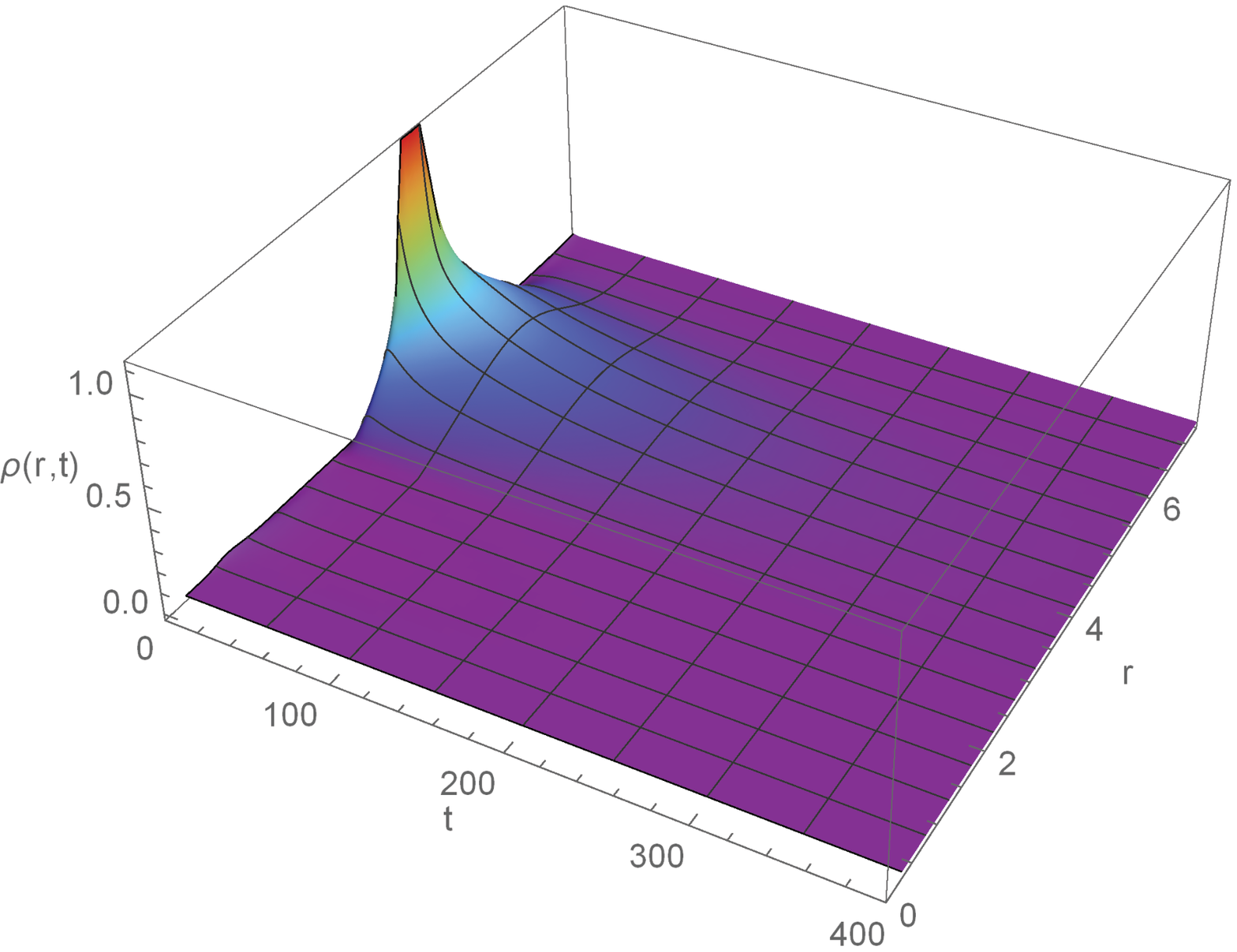}}
\subfigure[]{\label{a0.2p0.003s3}
\includegraphics[width=8cm,height=6cm]{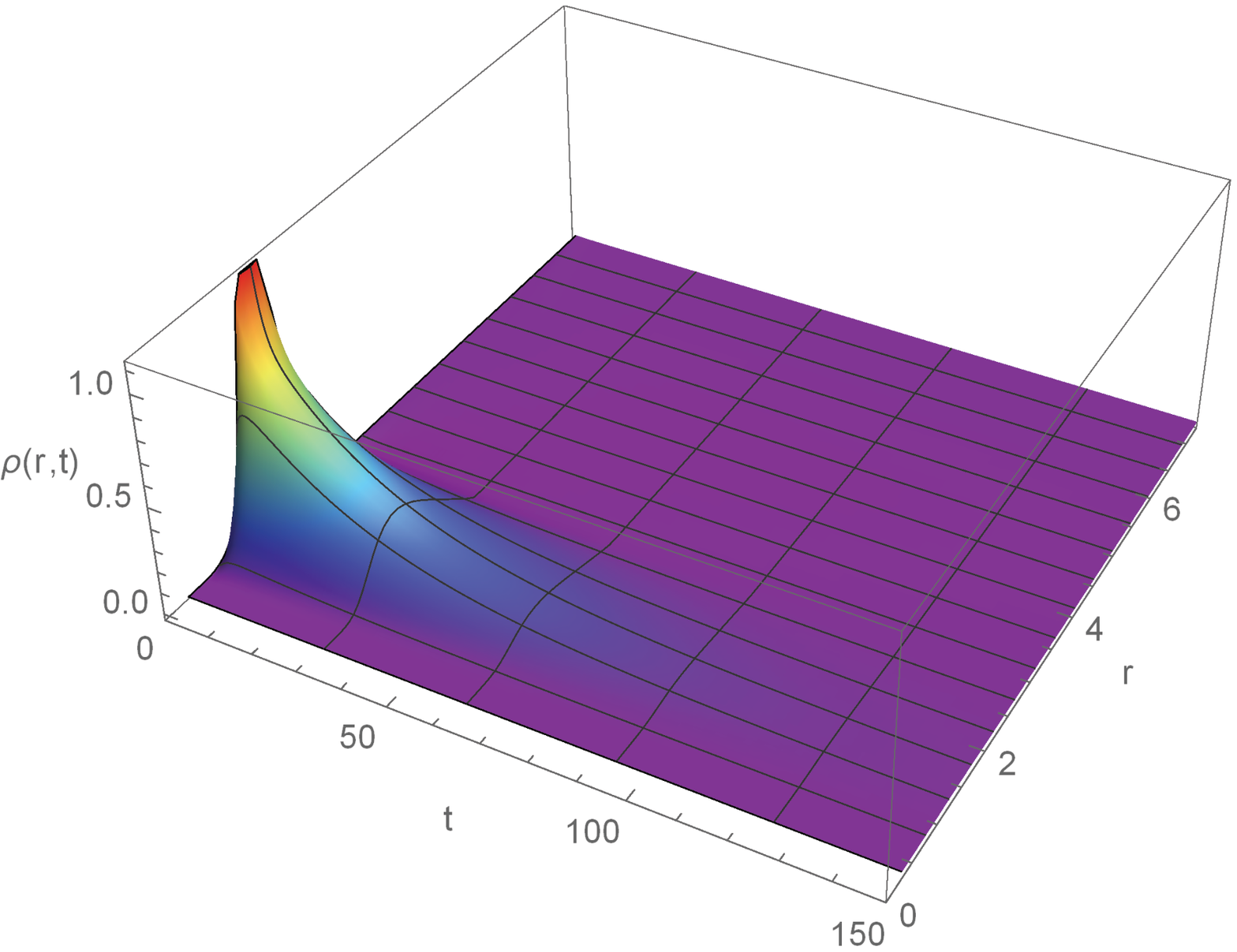}}}
\centerline{\subfigure[]{\label{a0p0.003l3}
\includegraphics[width=8cm,height=6cm]{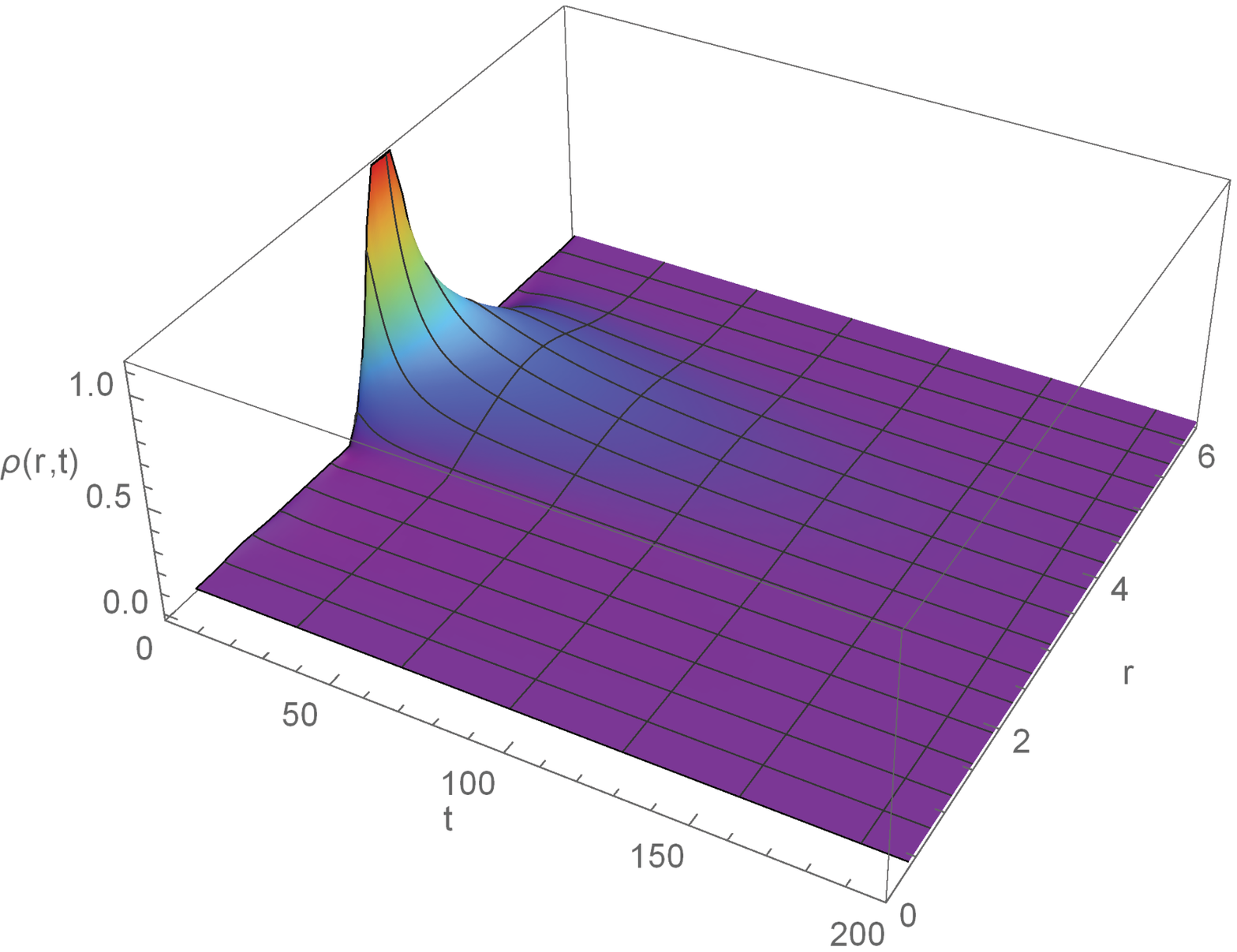}}
\subfigure[]{\label{a0p0.003s3}
\includegraphics[width=8cm,height=6cm]{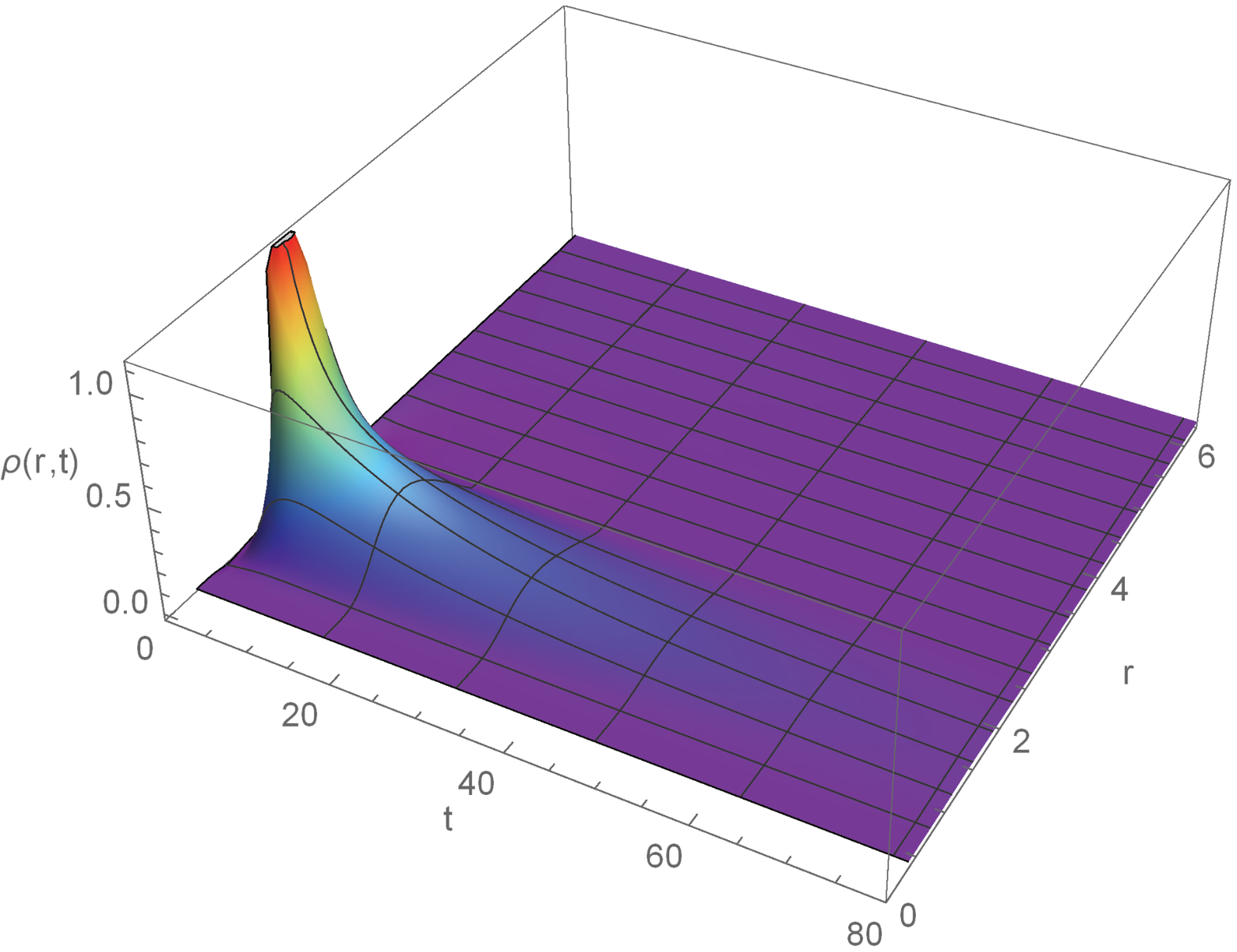}}}
 \caption{Time evolution of $\rho(r,t)$ in the first passage process of charged dilaton black hole with $b=1, q=1, P=0.003$. (a) $\alpha=0.4$ (b) $\alpha=0.4$  (c) $\alpha=0.2$  (d) $\alpha=0.2$ (e) $\alpha=0$ (f) $\alpha=0$. For the left three graphs, the initial condition is chosen as Gaussian wave pocket located at the large dilaton black hole state while for the right three graphs, the initial condition is chosen as Gaussian wave pocket located at the small dilaton black hole state.}
\label{fg3}
\end{figure}

Resolving the Fokker-Planck equation constrained by these two types of boundary conditions, we show the first passage process of charged dilaon black holes intuitively in Fig.\ref{fg3}. Similarly,  the initial condition is chosen as Gaussian wave pocket located at the large dilaton black hole state for the left three graphs while the initial condition is chosen as Gaussian wave pocket located at the small dilaton black hole state for the right two graphs. No matter what the initial black hole state is, the initial peaks decreases very quickly. We also compare the evolution for different choices of parameter $\alpha$. In the top, middle and bottom rows of Fig.\ref{fg3}, $\alpha$ is chosen respectively as $0.4,0.2,0$. As can be witnessed from Fig.\ref{fg3}, the initial peak decays more slowly with the increase of $\alpha$, consistent with the findings in Fig.\ref{fg1} and Fig.\ref{fg2}.

The time needed for the first passage process is denoted as the first passage time. Its distribution $F_p(t)$ can be obtained via~\cite{liran1}
\begin{equation}
F_p(t)=-\frac{d\Sigma(t)}{dt}.\label{12}
\end{equation}
where $\Sigma(t)$ is the sum of the probability that the black hole system having not finished a first passage by time $t$.

\begin{figure}[H]
\centerline{\subfigure[]{\label{a0.4p0.003l4}
\includegraphics[width=8cm,height=6cm]{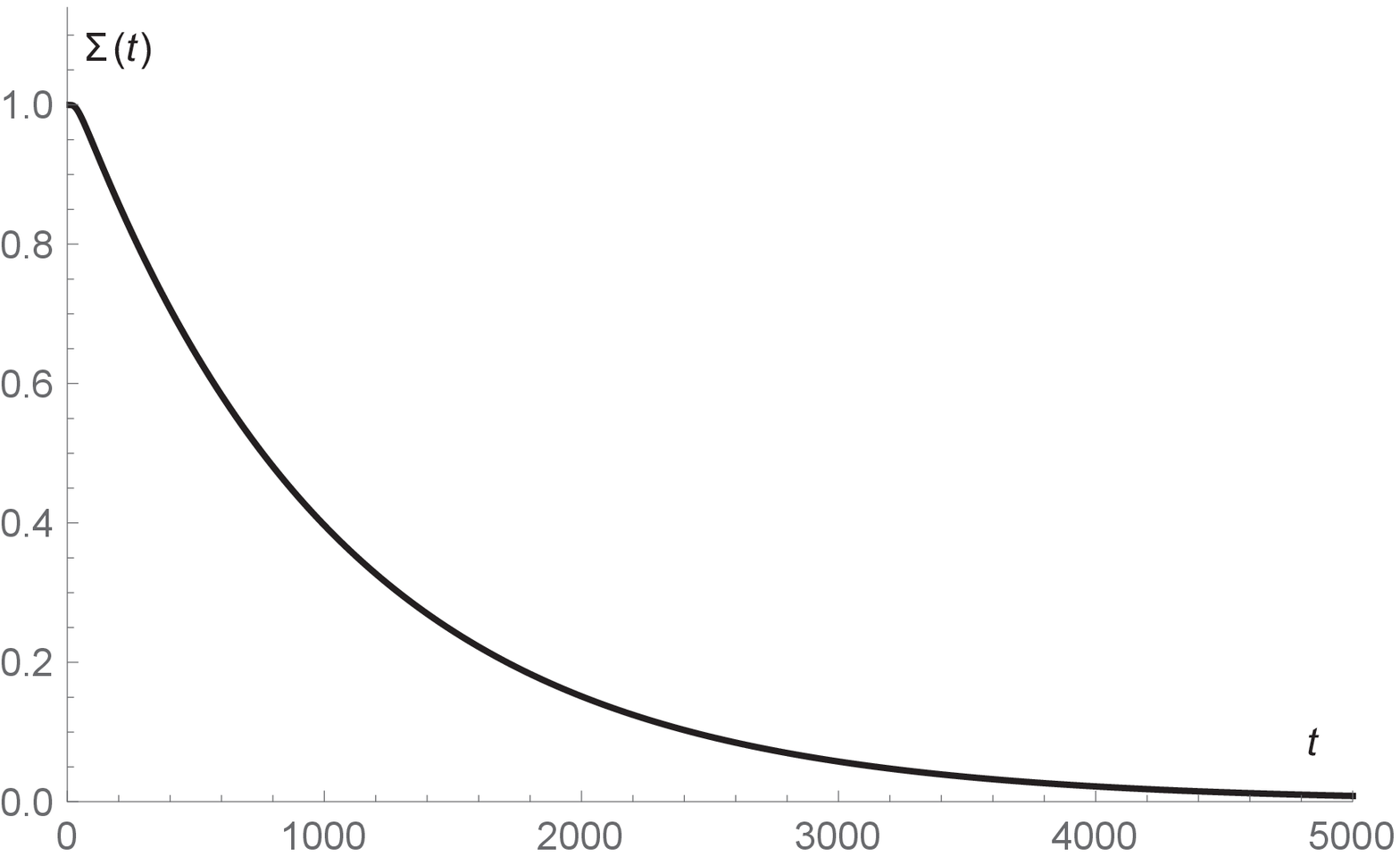}}
\subfigure[]{\label{a0.4p0.003s4}
\includegraphics[width=8cm,height=6cm]{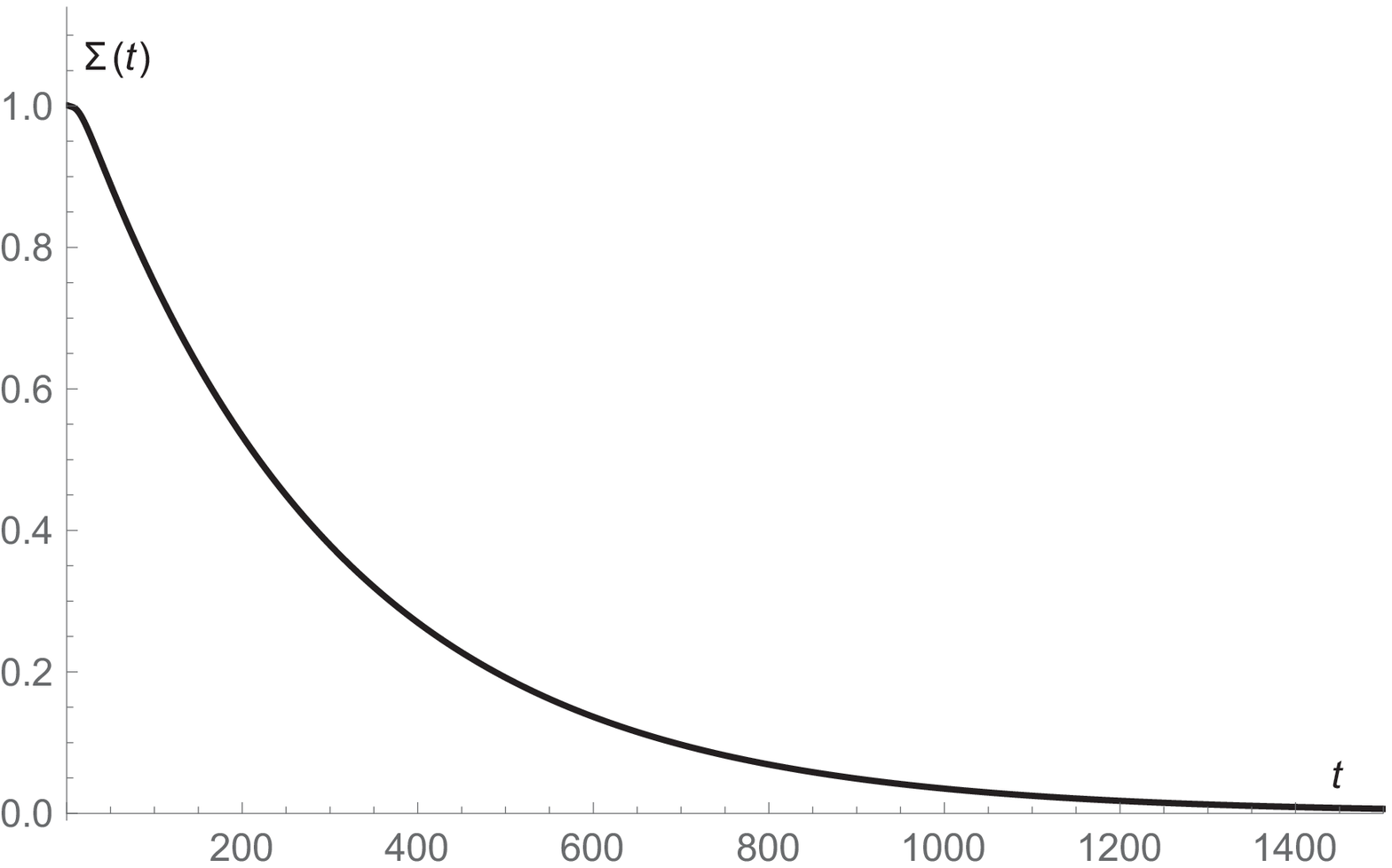}}}
\centerline{\subfigure[]{\label{a0.2p0.003l4}
\includegraphics[width=8cm,height=6cm]{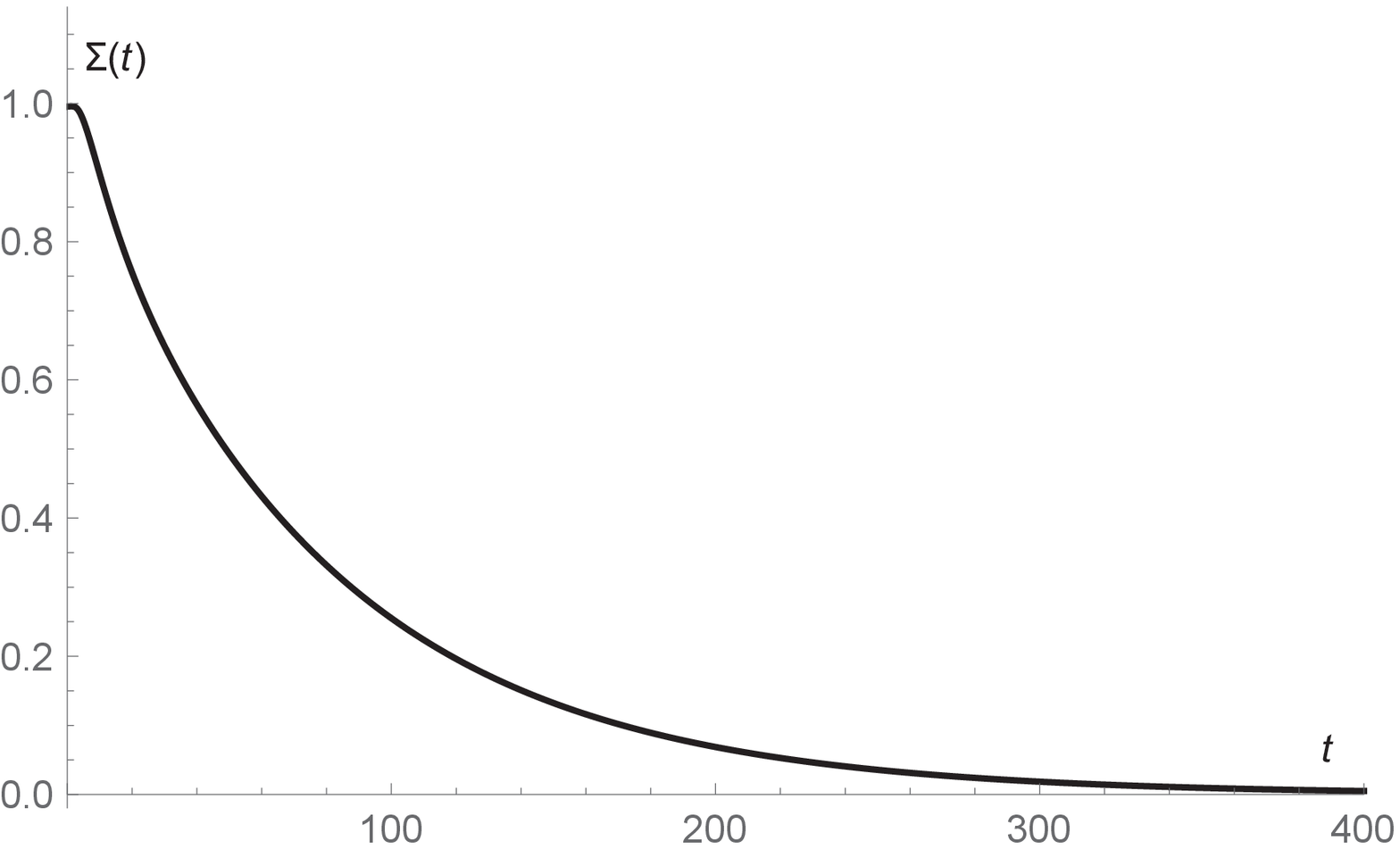}}
\subfigure[]{\label{a0.2p0.003s4}
\includegraphics[width=8cm,height=6cm]{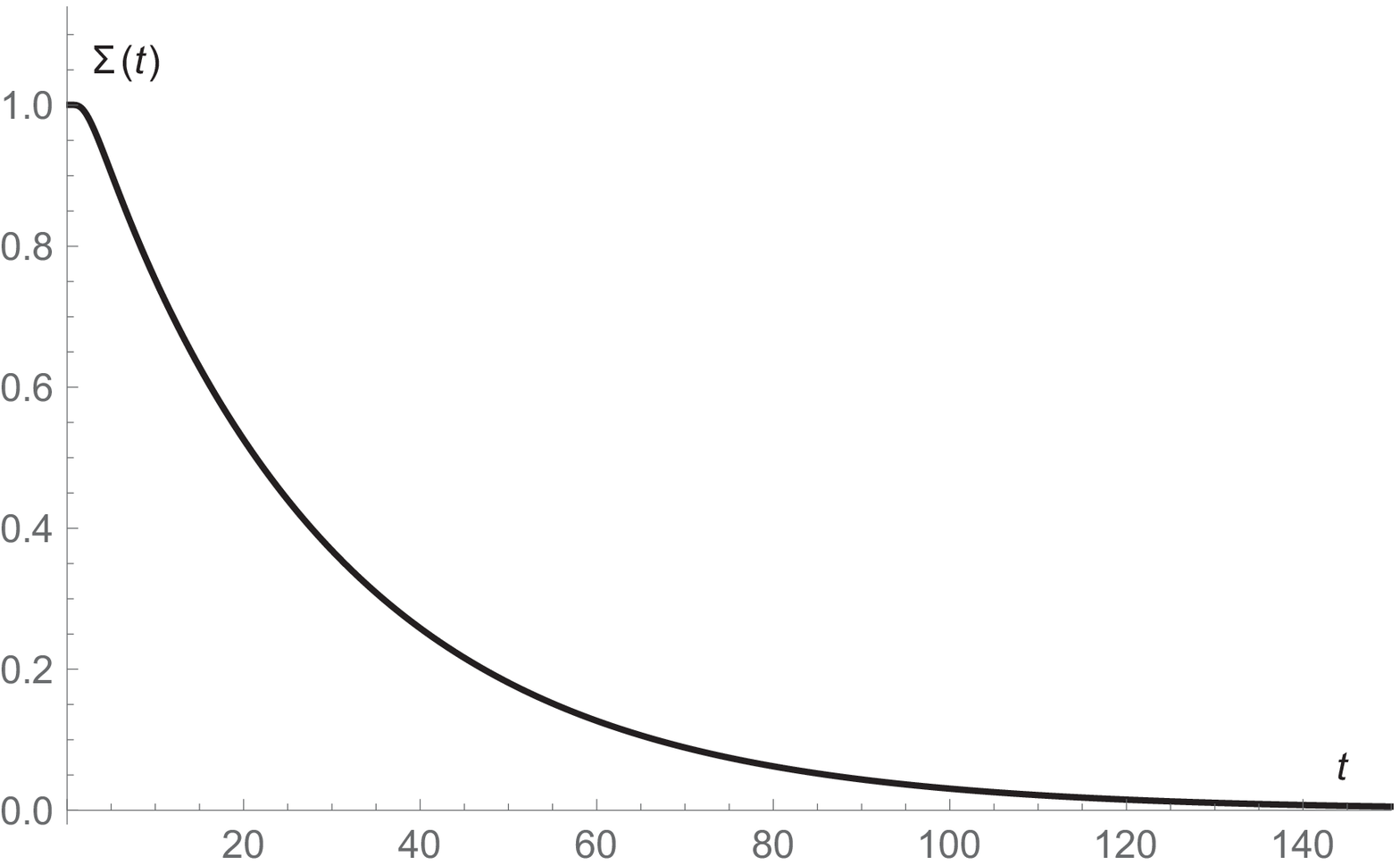}}}
\centerline{\subfigure[]{\label{a0p0.003l4}
\includegraphics[width=8cm,height=6cm]{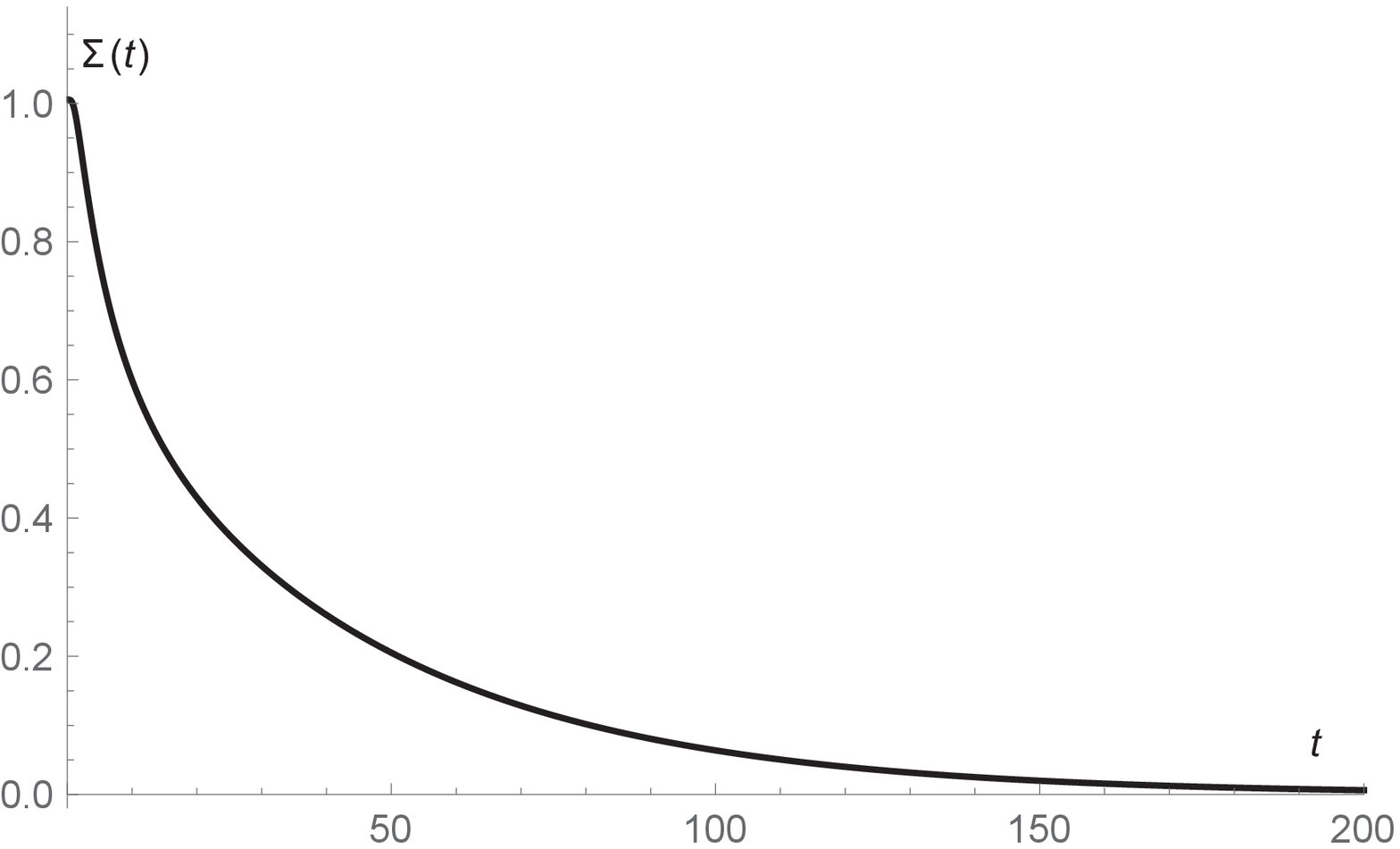}}
\subfigure[]{\label{a0p0.003s4}
\includegraphics[width=8cm,height=6cm]{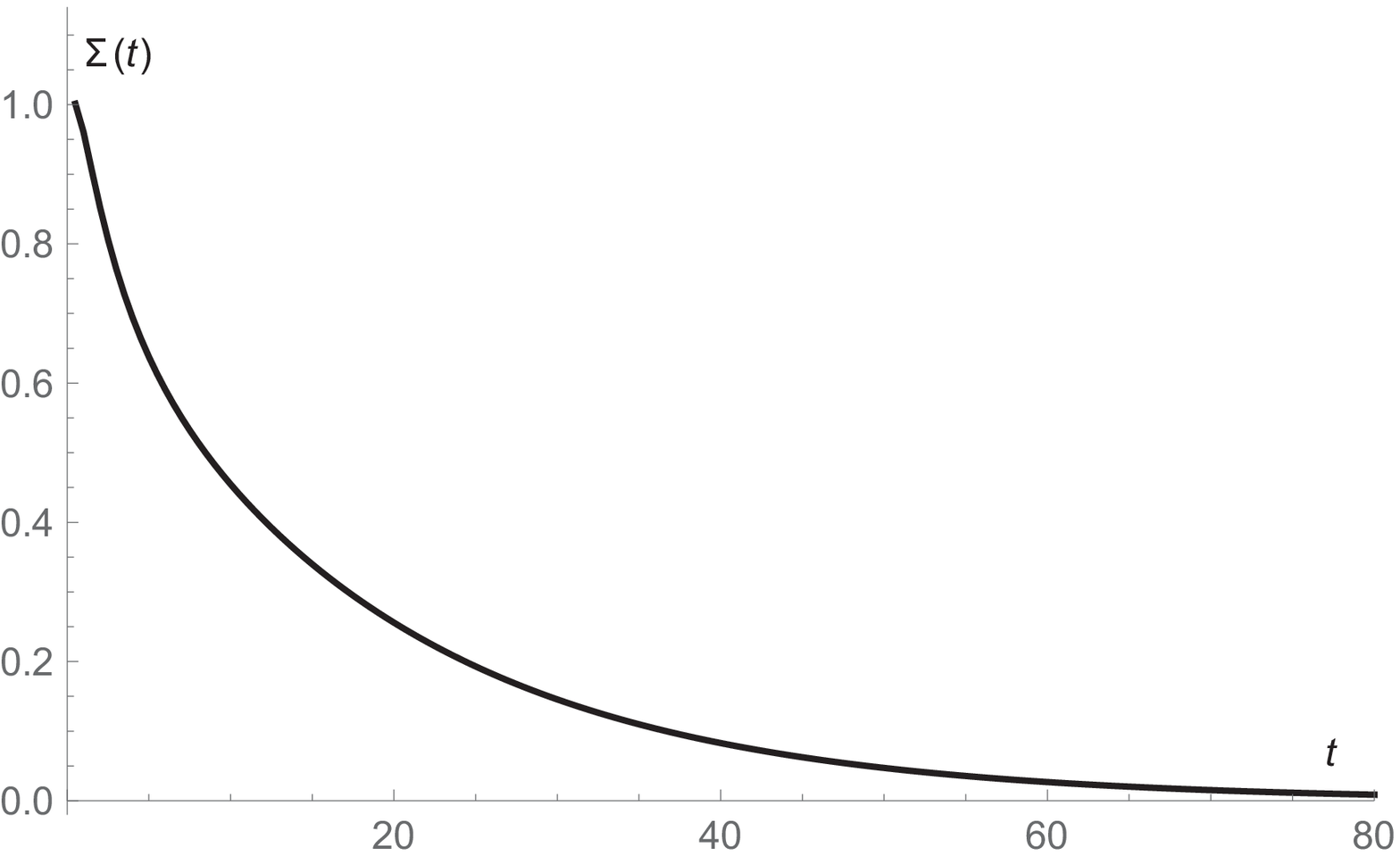}}}
 \caption{Time evolution of $\Sigma(t)$ for charged dilaton black hole with $b=1, q=1, P=0.003$. (a) $\alpha=0.4$ (b) $\alpha=0.4$  (c) $\alpha=0.2$  (d) $\alpha=0.2$ (e) $\alpha=0$ (f) $\alpha=0$. For the left three graphs, the initial condition is chosen as Gaussian wave pocket located at the large dilaton black hole state while for the right three graphs, the initial condition is chosen as Gaussian wave pocket located at the small dilaton black hole state.}
\label{fg4}
\end{figure}

Fig.\ref{fg4} shows the time evolution of $\Sigma(t)$. Irrespective of the initial state, $\Sigma(t)$ decays very fast. The effect of dilaton gravity can also be obviously witnessed. $\alpha$ is chosen respectively as $0.4,0.2,0$ for the top, middle and bottom rows of Fig.\ref{fg4}. With the increasing of $\alpha$,  the decay of $\Sigma(t)$ slows down.

$F_p(t)$ can also be derived via Eqs. (\ref{11}) and (\ref{12}) as~\cite{liran2}
\begin{equation}
F_p(t)=-D\frac{\partial \rho(r,t)}{\partial r}\Big|_{r_m},\label{14}
\end{equation}

\begin{figure}[H]
\centerline{\subfigure[]{\label{a0.4p0.003l5}
\includegraphics[width=8cm,height=6cm]{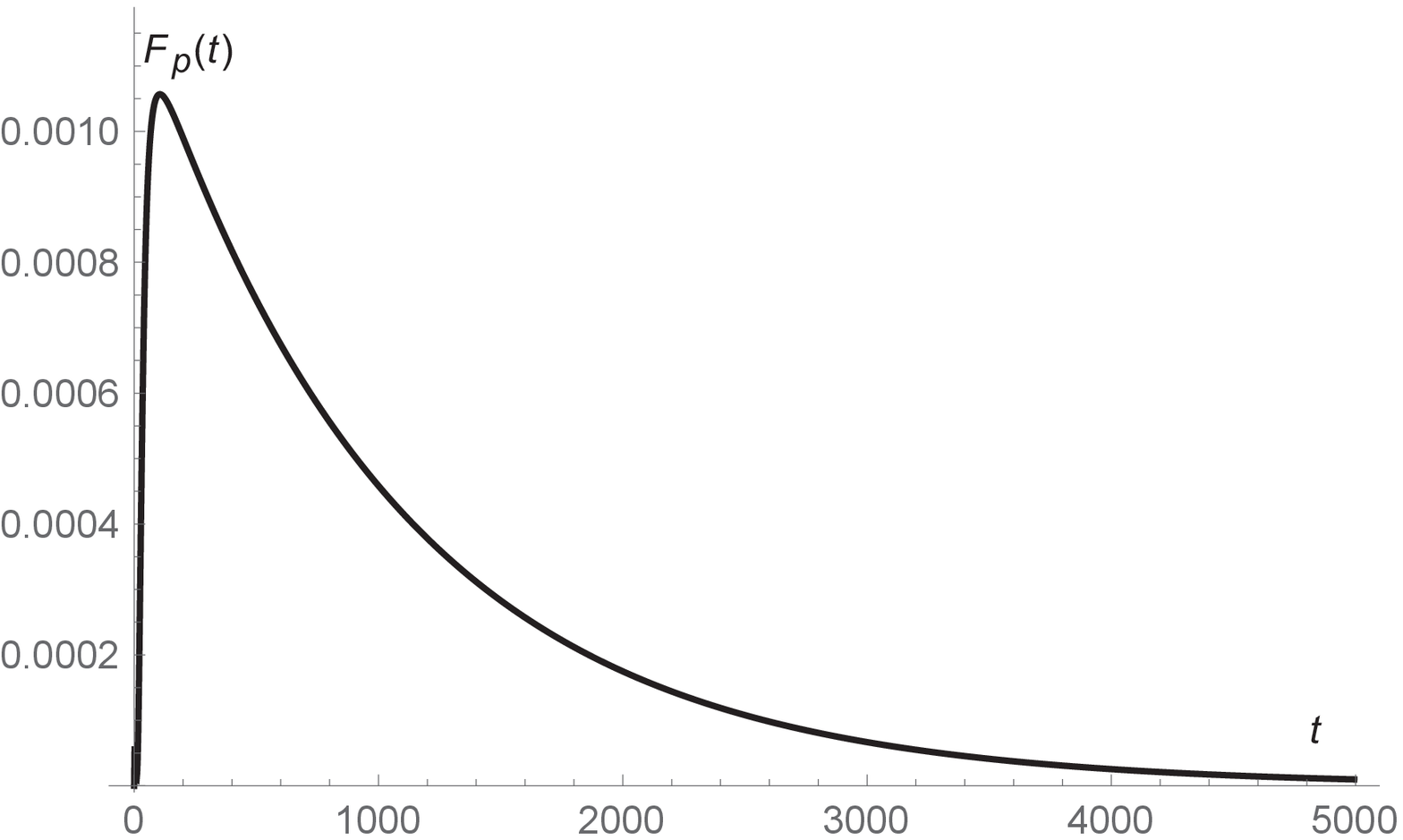}}
\subfigure[]{\label{a0.4p0.003s5}
\includegraphics[width=8cm,height=6cm]{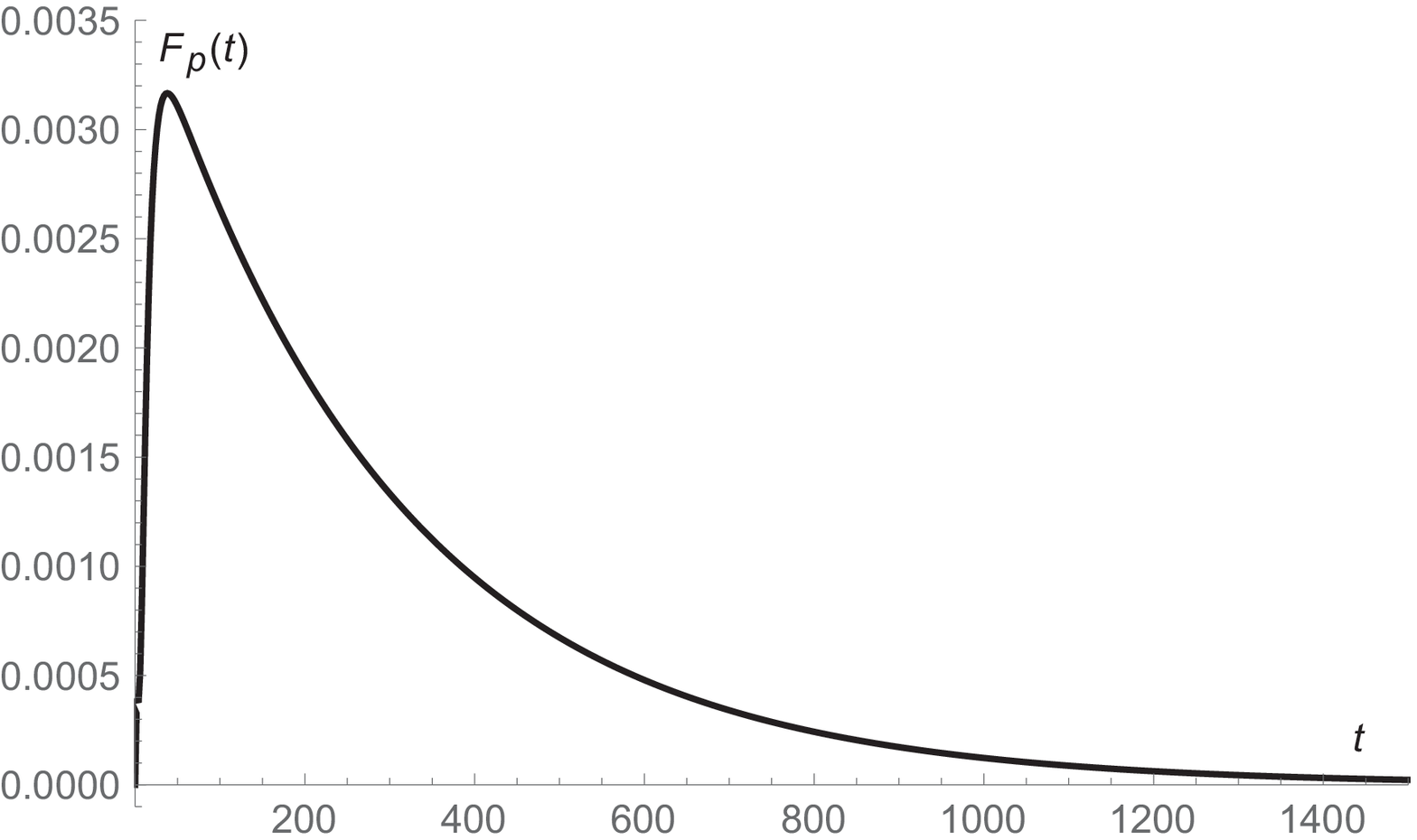}}}
\centerline{\subfigure[]{\label{a0.2p0.003l5}
\includegraphics[width=8cm,height=6cm]{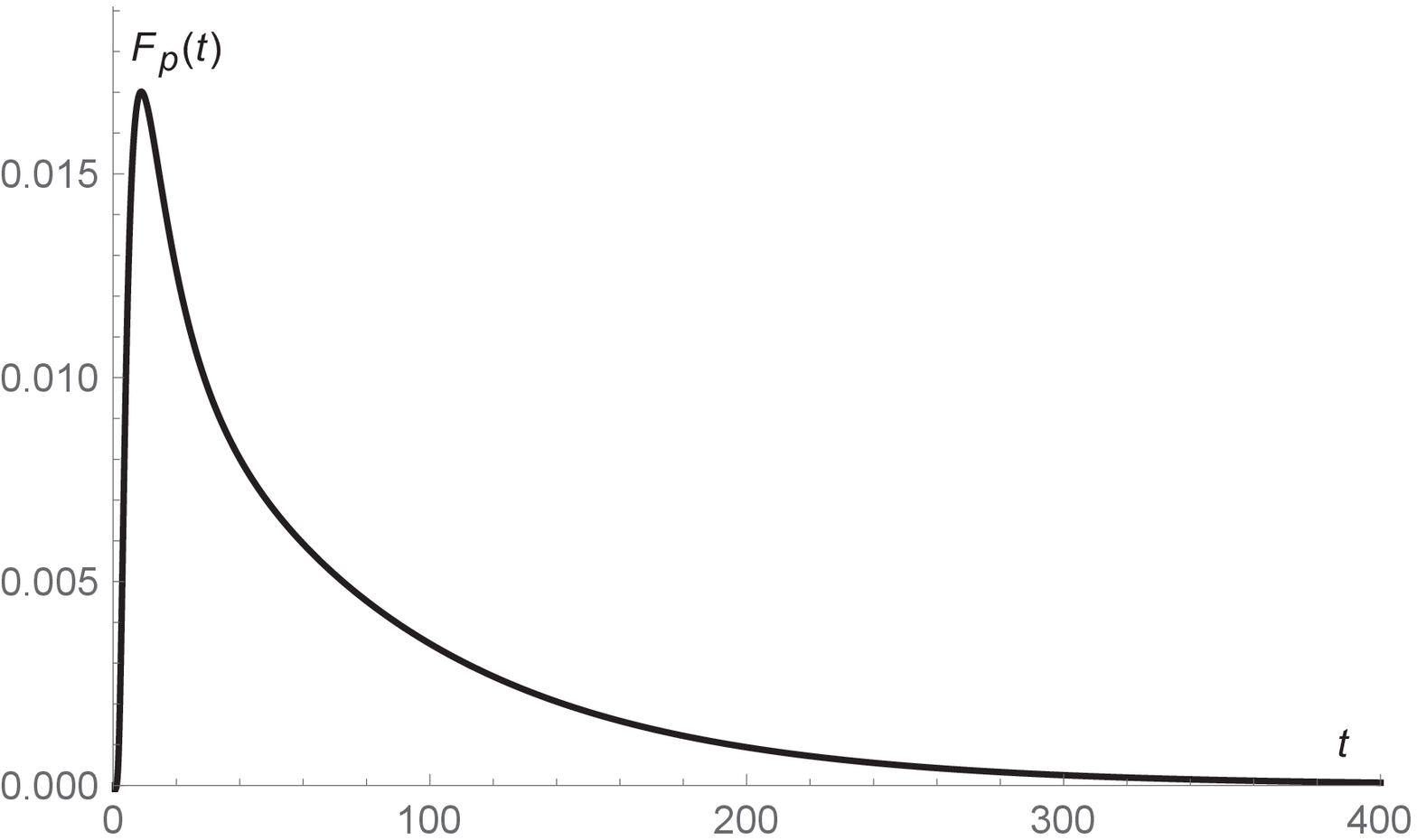}}
\subfigure[]{\label{a0.2p0.003s5}
\includegraphics[width=8cm,height=6cm]{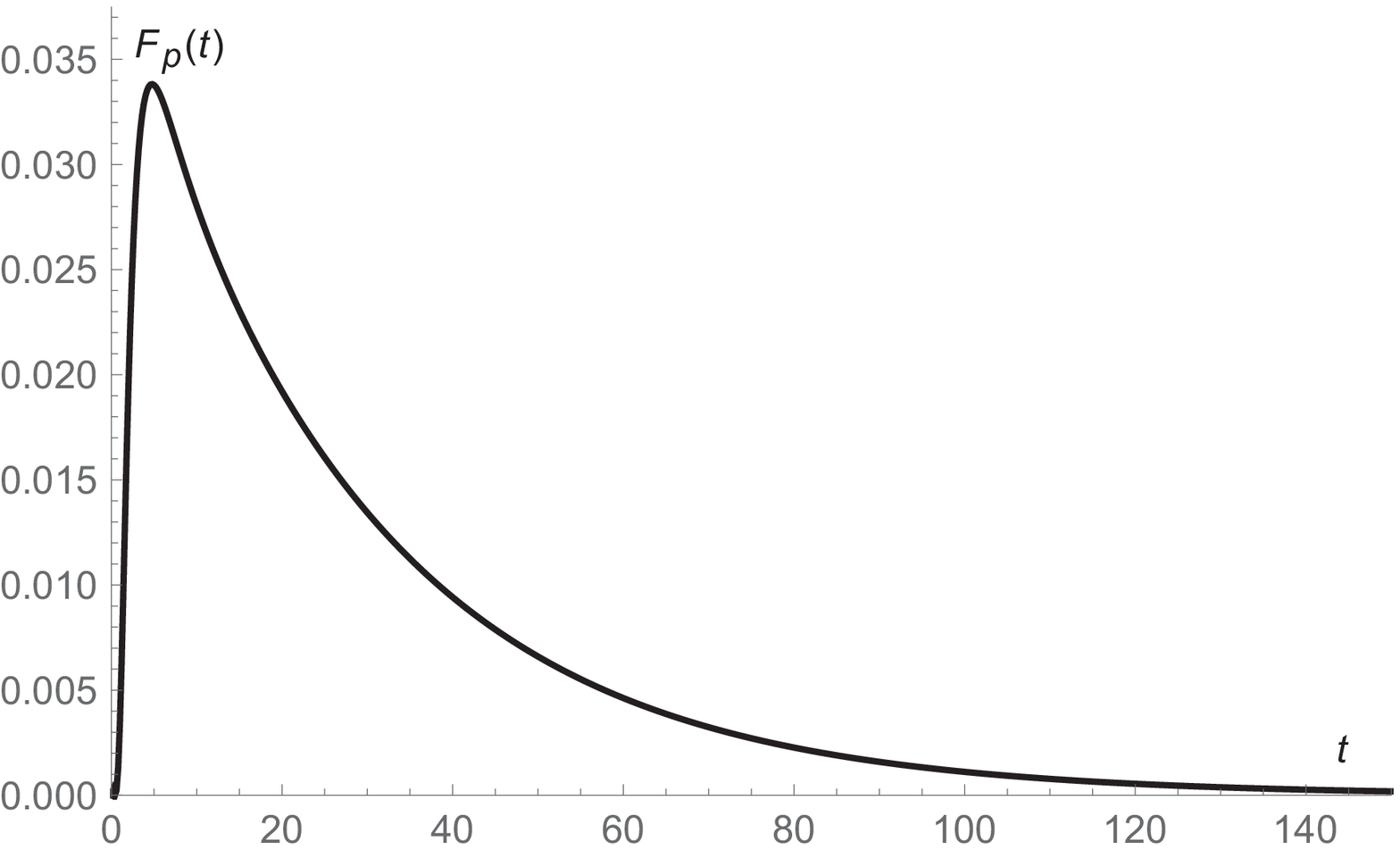}}}
\centerline{\subfigure[]{\label{a0p0.003l5}
\includegraphics[width=8cm,height=6cm]{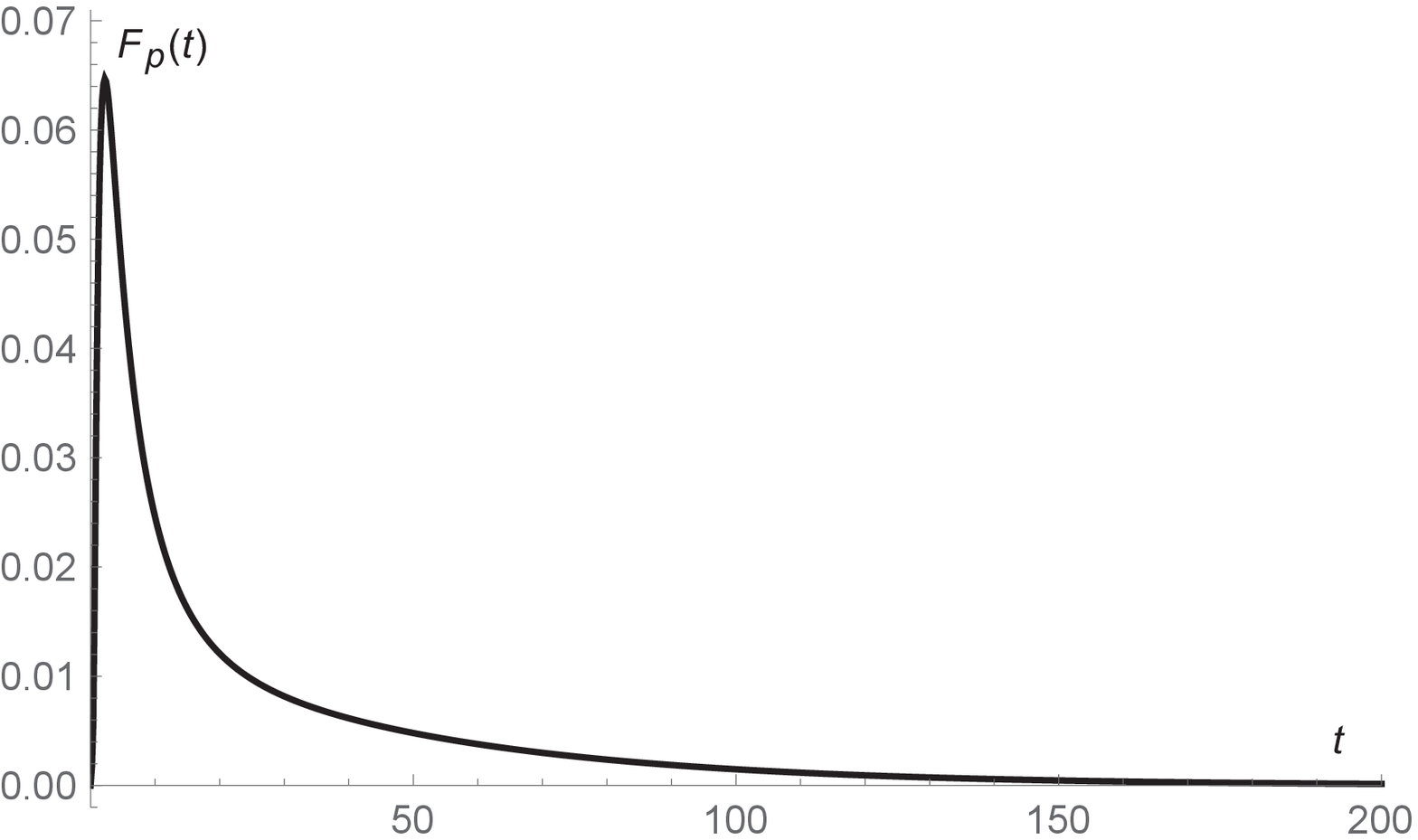}}
\subfigure[]{\label{a0p0.003s5}
\includegraphics[width=8cm,height=6cm]{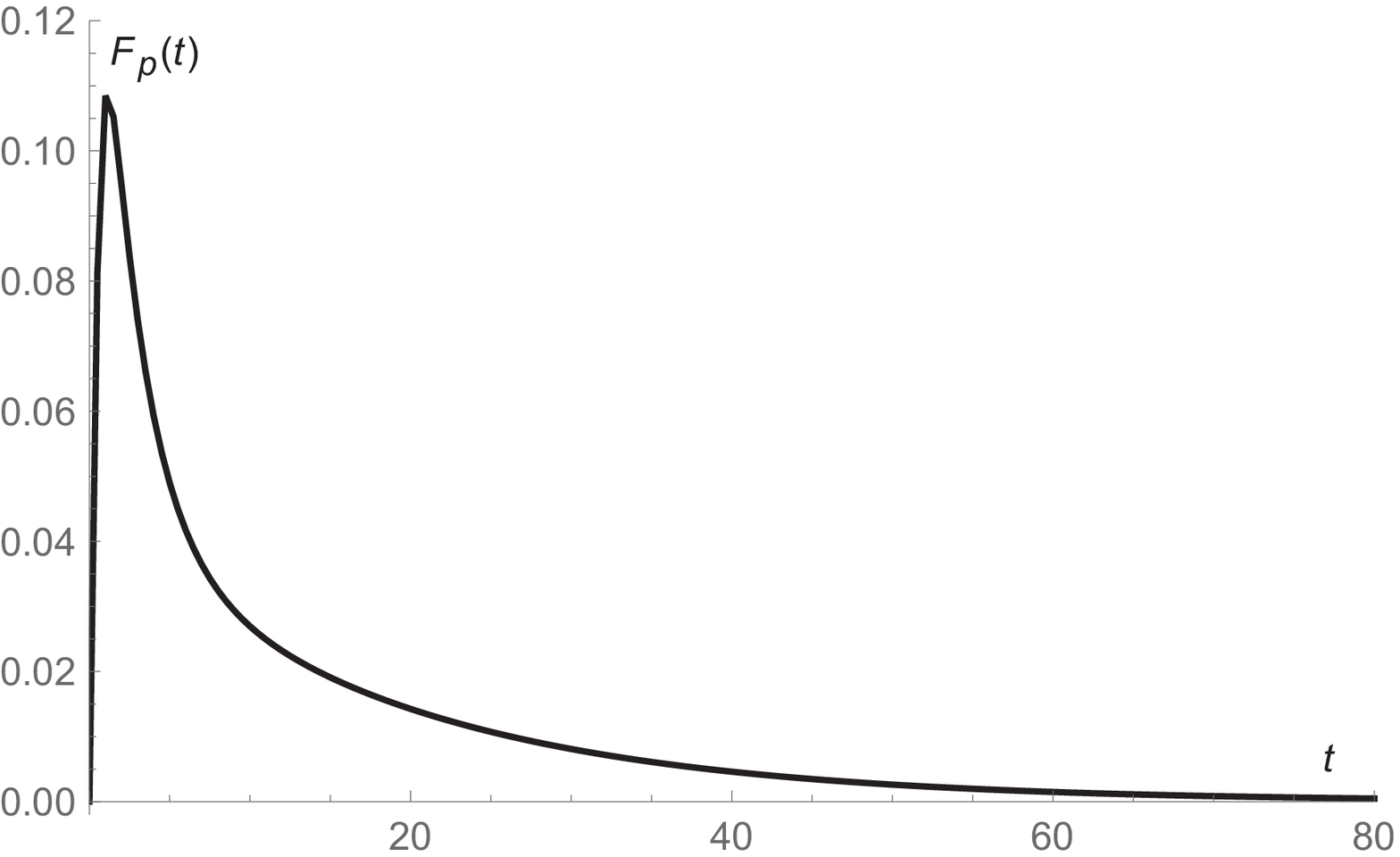}}}
 \caption{$F_p(t)$ for charged dilaton black hole with $b=1, q=1, P=0.003$. (a) $\alpha=0.4$ (b) $\alpha=0.4$  (c) $\alpha=0.2$  (d) $\alpha=0.2$ (e) $\alpha=0$ (f) $\alpha=0$. For the left three graphs, the initial condition is chosen as Gaussian wave pocket located at the large dilaton black hole state while for the right three graphs, the initial condition is chosen as Gaussian wave pocket located at the small dilaton black hole state.}
\label{fg5}
\end{figure}

The distribution of first passage time is depicted in Fig.\ref{fg5}. It is shown that there exists one single peak in all the graphs of $F_p(t)$. This observation suggests that the first passage process takes place very fast. And the time corresponding to the peak increases with the parameter $\alpha$.

%%%%%%%%%%%%%%%%%%%%%%%%%%%%%%%%%%%%%%%%%%%%%%%%%%%%%%%%%%%%%%%%%%%%%%%%%%%%%
%%%%%%%%%%%%%%%%%%%%%%%%%%%%%%%%%%%%%%%%%%%%%%%%%%%%%%%%%%%%%%%%%%%%%%%%%%%%%%%%

%%%%%%%%%%%%%%%%%%%%%%%%%%%%%%%%%%%%%%%%%%%%%%%%%%%%%%%%%%%%%%%%%%%%%%%%%%%%%
%%%%%%%%%%%%%%%%%%%%%%%%%%%%%%%%%%%%%%%%%%%%%%%%%%%%%%%%%%%%%%%%%%%%%%%%%%%%%%%%

%%%%%%%%%%%%%%%%%%%%%%%%%%%%%%%%%%%%%%%%%%%%%%%%%%%%%%%%%%%%%%%%%%%%%%%%%%%%%
%%%%%%%%%%%%%%%%%%%%%%%%%%%%%%%%%%%%%%%%%%%%%%%%%%%%%%%%%%%%%%%%%%%%%%%%%%%%%%%%

\section{Conclusions}
\label{Sec5}
We probe the effect of dilaon gravity on the dynamic phase transition of black holes by investigating the case of charged dilaton black holes. Specifically, we consider the the variation of the parameter $\alpha$ describing the strength of coupling between the electromagnetic field and the dilaton field.

We introduce the Gibbs free energy landscape and calculate the corresponding $G_L$ of the dilaton black hole. Based on $G_L$, we can numerically solve the Fokker-Planck equation. To probe the probabilistic evolution of dilaton black holes, we first consider the reflecting boundary condition. When the initial condition is chosen as the large dilaton black hole, the initial peak of $\rho(r,t)$ corresponding to the radius of the large dilaton black hole is decreasing with time while another peak corresponding to the radius of the small dilaton black hole is booming with time. Finally, these two peaks reach a stationary distribution where $\rho(r_l,t)$ and $\rho(r_s,t)$ attain the same value. This process clearly illustrates how the initial large dilaton black hole evolves dynamically into the small dilaton black hole. Similar analysis hold for the case that the initial condition is chosen as the small dilaton black hole. The effect of dilaton gravity on the probabilistic evolution of dilaton black holes is quite obvious. Firstly, the horizon radius difference between the large dilaton black hole and small dilaton black hole increases with the parameter $\alpha$. Secondly, with the increasing of $\alpha$, the system needs much more time to achieve a stationary distribution. Note that these two observations do not depend on the initial condition. Thirdly, the value which $\rho(r_l,t)$ and $\rho(r_s,t)$ attain varies with the parameter $\alpha$ ($0.229$ for the case of $\alpha=0.4$, $0.250$ for the case of $\alpha=0.2$ and $0.297$ for the case of $\alpha=0$).

To further probe the dynamic phase transition of dilaton black holes, we consider first passage process which describes how the initial black hole state approach the intermediate transition state for the first time. Besides the reflecting boundary condition, absorbing boundary condition should also be imposed here. Resolving the Fokker-Planck equation constrained by these two types of boundary conditions, we show the first passage process of charged dilaon black holes intuitively. No matter what the initial black hole state is, the initial peak decreases very quickly. We compare the evolution for different choices of $\alpha$. The initial peak decays more slowly with the increase of $\alpha$. We also study the distribution of the first passage time $F_p(t)$ and the sum of the probability $\Sigma(t)$ that the black hole system having not finished a first passage by time $t$. Irrespective of the initial state, $\Sigma(t)$ decays very fast. The effect of dilaton gravity can also be obviously witnessed. With the increasing of $\alpha$,  the decay of $\Sigma(t)$ slows down. There exists one single peak in all the graphs of $F_p(t)$, suggesting that the first passage process takes place very fast. And the time corresponding to the peak is found to increase with the parameter $\alpha$.

To conclude, the dilaton field slows down the dynamic phase transition process between the large black hole and the small black hole. It would also be interesting to probe the effect of dilaton field on the dynamic process of novel phase transition disclosed in Ref. \cite{Sheykhi16}.

 \section*{Acknowledgements}

 Shan-Quan Lan is supported by National Natural Science Foundation of China (Grant No.12005088) and Lingnan Normal University Project with Grant Nos.YL20200203, ZL1930. The authors are also in part supported by Guangdong Basic and Applied Basic Research Foundation, China (Grant No.2021A1515010246).

\end{document}